\pdfoutput=1 
\documentclass{agujournal2019}
\usepackage{url} 
\usepackage[footnotes]{trackchanges} 
\usepackage{soul}
\linenumbers

\usepackage{amsmath,amssymb,amsfonts}
\usepackage{amsthm}
\usepackage{bm}
\usepackage[capitalise]{cleveref} 
\usepackage{enumerate,enumitem}
\usepackage[linesnumbered,ruled,vlined,algo2e]{algorithm2e}
\usepackage{hyphenat}
\journalname{Journal of Advances in Modeling Earth Systems (JAMES)}

\newcommand{\X}{\mathcal{X}}

\crefname{algocf}{algorithm}{Algs.}
\Crefname{algocf}{Algorithm}{Algorithms}
\addeditor{lmy}

\begin{document}
\nolinenumbers

%
%


\title{Overcoming set imbalance in data driven parameterization: A case study of gravity wave momentum transport}

%
%




\authors{L. Minah Yang \affil{1}, Edwin P. Gerber \affil{1}}


\affiliation{1}{Center for Atmosphere Ocean Science, Courant Institute of
Mathematical Sciences, New York University, New York, New
York, USA.}




\correspondingauthor{L. Minah Yang}{minah.yang@nyu.edu}



\begin{keypoints}
\item Unresolved geophysical processes often exhibit long tail distributions, which leads to imbalanced datasets for data-driven parameterizations.
\item Two strategies to overcome data imbalance are presented, where either the sampling or loss function is modified to better capture the tails.
\item Proof of concept is demonstrated by using a wind range metric to improve a machine learning emulator of a physics based gravity wave parameterization.
\end{keypoints}

%
%

%
%


\begin{abstract}
Machine learning for the parameterization of subgrid-scale processes in climate models has been widely researched and adopted in a few models.
A key challenge in developing data-driven parameterization schemes is how to properly represent rare, but important events that occur in geoscience datasets. 
We investigate and develop strategies to reduce errors caused by insufficient sampling in the rare data regime, under constraints of no new data and no further expansion of model complexity.
Resampling and importance weighting strategies are constructed with user defined parameters that systematically vary the sampling/weighting rates in a linear fashion
and curb too much oversampling.
Applying this new method to a case study of gravity wave momentum transport reveals that the resampling strategy can successfully improve errors in the rare regime at little to no loss in accuracy overall in the dataset.
The success of the strategy, however, depends on the complexity of the model. 
More complex models can overfit the tails of the distribution when using non-optimal parameters of the resampling strategy.

\end{abstract}

\section*{Plain Language Summary}
Subgrid-scale parameterizations are a part of climate models that represent effects of processes that cannot be directly modelled. 
In recent years, there have been many efforts to improve upon these parameterizations by applying machine learning techniques. 
Since these methods rely heavily on the dataset they are learning from, it is important to consider the frequency at which important events occur within the dataset because they are adept at learning frequent events at high accuracy but are prone to learning rare but important events at low accuracy.
To remedy this \emph{data imbalance} problem, we developed a resampling methodology that can be easily adjusted by tuning just two parameters. 
We find that a right combination of those parameters can improve the accuracy of an ML model at the rare event regime while keeping the accuracy high in the frequent regime. 
However, a ``wrong'' combination can actually increase the errors at the rare event regime by overfitting to that regime.

%
%

\section{Introduction} \label{sec:intro}
Machine learning techniques have been used to develop data driven parameterization of un- or under-resolved processes in climate models, including a comprehensive representation of all missing terms, either at once \cite{brenowitz_spatially_2019} or separately \cite{yuval_use_2021}, or specific processes, including gravity wave momentum transport \cite{chantry_machine_2021,espinosa_machine_2022} and radiative transfer \cite{ukkonen_exploring_2022}.
None of these attempts yielded a perfect sub-grid scale model, begging a general question: what can one do to improve a given data-driven parameterization?
As these processes, and geoscience datasets more generally, are often high-dimensional and exhibit long-tailed distributions, a common problem is to properly learn rare and extreme events.  
This is particularly problematic if these extreme events have an outsized impact on the climate, or become more prevalent in a changing climate.
How can we capture important but rare events from the tail of the distribution as best as possible given the dataset available to us? 
This is a data imbalance problem, and we propose strategies to combat it in this paper.

Set imbalance is a common challenge in machine learning (ML).
In binary classification, the imbalanced dataset problem refers to a skewed distribution of the two target classes in a dataset.
A naive learning algorithm will inherit an asymmetric class representation in the dataset, and will typically produce classifiers that predict the minority class with lower accuracy than it does for the majority class.
These biased classifiers prove even more problematic when the minority class holds more importance or utility. 
As this combination of challenges is ubiquitous in real datasets, many methods that curb and minimize biases that stem from imbalanced datasets have been developed, as reviewed by \citeA{he_learning_2009} and \citeA{krawczyk_learning_2016}. 

Data imbalance poses difficulties for ML tasks outside of binary classification.
While it is straightforward to extend methods for treating imbalanced datasets for binary to multi-class classification, it  has proven more difficult to extend this for regression tasks.
Here, one seeks to learn a function $g$ from a set of inputs $\vec{x}$ to outputs $\vec{y}$ where the example pairs $(\vec{x},\vec{y})$ is unevenly distributed.
As with the classification problem, the task is particularly hard if we care especially about the behavior of $g$ for rare pairs of $(\vec{x},\vec{y})$.

In this paper, we explore systematic methods for overcoming data imbalance in regression tasks,  illustrating them with a case study of data driven parameterization gravity wave (GW) momentum transport.  
Gravity waves play an important role in forcing the large scale atmospheric circulation, but their small scale makes them challenging to properly represent directly.  
We seek a function $g$ that maps vertical profiles of the resolved wind, temperature, and GW source information within a column of an atmospheric model: $\vec{x}$, to the profiles of the grid scale momentum tendency by unresolved gravity waves associated with this large scale environment: $\vec{y}$.  
We assume limited resources, in that one cannot simply increase the size of the dataset or complexity of our model $g$ to overcome the problem: the goal is to work with the data and model one has on hand.

First steps have been taken towards deriving data-driven schemes for GWs by exploring how well machine learning approaches can emulate existing, physics based parameterizations \cite{chantry_machine_2021,sun_quantifying_2023}.
Both studies found that data imbalance was challenging, particularly for capturing the momentum forcing by gravity wave excited by orography. 
Not only are most grid cells of a GCM flat, but even where there is topography, the waves themselves are highly intermittent.  
Here, we will focus on non-orographic waves, but the method is general and an ad hoc version of it was used by \citeA{sun_quantifying_2023} to emulate an orographic paramterization.  
More specifically, we build on the work of \citeA{espinosa_machine_2022}, who emulated a physics-based GW parameterization (GWP) scheme \cite{Alexander1999} hereafter referred to as AD99, with a deep neural network (DNN) architecture called WaveNet. 
We continue this investigation to illustrate our approach for improving a generic ML methodology.  Exploring our method in the context of emulation also allows us to explore the ability of a scheme to generalize to different climates. 

The strategy involves two distinct steps.  
First, one must identify the data imbalance.  
This requires ``domain knowledge" of the problem, to identify key metric(s) that quantify rare cases where errors in the data-driven scheme limit its effectiveness.  
As detailed in Section \ref{sec:background}, we establish a wind range metric to identify rare cases where WaveNet enmulator systematically fails.  
On top of being rare, these are cases where the physics of AD99 scheme become more non-local, and so more challenging to learn.

Once the data imbalance is identified, the second  step is to treat it during model training and implementation, as detailed in Section \ref{sec:methods}.  
We illustrate two strategies at the learning stage, either to modify the sampling of training examples so that rarer cases are better represented from the start, or to leave the distribution as it is, but adjust the loss function to more strongly penalize mistakes on the rare cases.  To construct a principled method for this rebalancing, we borrow a concept from histogram equalization: a linear interpolation of the original distribution to a more uniform distribution parameterized by a scalar $t$ which can be varied from 0, where no change is made, to 1, where the distribution is made completely uniform.  The goal is to improve representation of the rare cases without losing skill on the central part of the distribution or overfitting the data in the tails, and the parameter $t$ allows one to calibrate the degree of rebalancing.   

These strategies assume that the ML model has enough complexity to learn the complex nonlinear behavior described by physics of $g$, but the data imbalance enables the model to ignore rare samples and predominantly learn from the typical samples.  
As we'll show in \cref{sssec:overfit}, overfitting can occur when the ML method is too complex with respect to the amount of training data available.
In addition to improving the training of an ML scheme, one can  mitigate data imbalance by applying a bias correction at the inference stage.
This involves computing the mean bias of the ML model as a function of the relevant metric (the wind range in our case study of GWP emulation), and subtracting the bias from the output.
The remainder of the paper is structured as follows.
\Cref{sec:background} illustrates how we identified  data imbalance,
\Cref{sec:methods} details modified training and bias removal methods to overcome this imbalance.
Our case study is presented in \cref{sec:case_study}. 
To demonstrate the generality of the method, we also introduce an alternative ML strategy, an Encoder-Dense-Decorder (EDD). 
We use our approach to improve both WaveNet and EDD.
Furthermore, we illustrate how our approach can fail when the complexity of the ML method exceeds the data available, leading to overfitting. 
\Cref{sec:conclusion} concludes our study and outlines possible future directions for this research.

\section{Identifying data imbalance}
\label{sec:background}

A first step towards improving a data-driven parameterization -- or more generally, any data-driven task -- is to identify potential imbalances in the training set.  
This process requires detailed knowledge of the application, as one is searching for metrics to quantify rare cases that are important for the performance of the task.  
The process is straightforward in low dimensional data sets, i.e., if one needs to differentiate cats from dogs, are the animals evenly distributed in the example data, but quickly becomes difficult in high dimensional datasets.  
Here, we illustrate an example where the input data has 83 dimensions, but we seek one particular dimension that clearly identifies rare, but important, samples that need to be learned.    

Our goal is to improve a data-driven emulator of the single column AD99 gravity wave parameterization, as implemented in the Model of an idealized Moist Atmosphere, MiMA \cite{garfinkel2020building}, following the work of \citeA{espinosa_machine_2022}.  
We direct the reader to \citeA{Alexander1999} for details on the parameterization and \citeA{espinosa_machine_2022} and \citeA{garfinkel2020building} for details on the atmospheric model, but briefly review the most salient points here.

As in \citeA{espinosa_machine_2022}, we use an integration of MiMA at triangular truncation T42 resolution (corresponding to a $\approx 3^{\circ}$ grid) with model parameters configured to produce a realistic representation of  northern hemisphere climate by \citeA{garfinkel2020building}. 
The model is integrated for 60 years, and after discarding the first 20 years' data as spin-up, we use years 21-30 for the training and years 56-60 for the validation set. 
Output from the model is saved 4 times a day, yielding over 1.1$\times 10^9$ samples, where each sample consists of vertical profiles of winds and temperature (the inputs), one for each column on a 128$\times$64 longitude-latitude grid, and the parameterized gravity wave tendency as the output.
For simplicity, we focus only on the zonal (East-West) gravity wave tendencies. 

AD99 is a multi-wave GW parameterization that adheres closely to the scheme established by \cite{lindzen_turbulence_1981}, which assumes the conservation of wave action flux and wave-mean flow interactions under linear theory. 
The scheme determines GW momentum transport by launching a spectrum of non-interacting, monochromatic waves. 
Thermodynamic breaking criteria determine when each wave breaks and deposits its momentum into the mean flow: waves tend to break when they near a critical level, where the speed of the large scale winds equals that of the GW, or when their amplitude becomes sufficiently large to overturn. 
This latter criteria is favored at upper levels where density decays.  Additional criteria account for waves that would be filtered out at the source level (the nominal tropopause) or reflected downward.
Important for our application, momentum carried by waves that do not break before reaching the model top are deposited in the upper levels of the column, thereby preventing a leak of momentum through the model top \cite{shaw2009sensitivity}.
A key simplification of the scheme is that the source spectrum is only a function of latitude, meant to capture a simple background of waves generated by convection, frontegenis, and orography.

\begin{figure}
  \includegraphics[width=\textwidth]{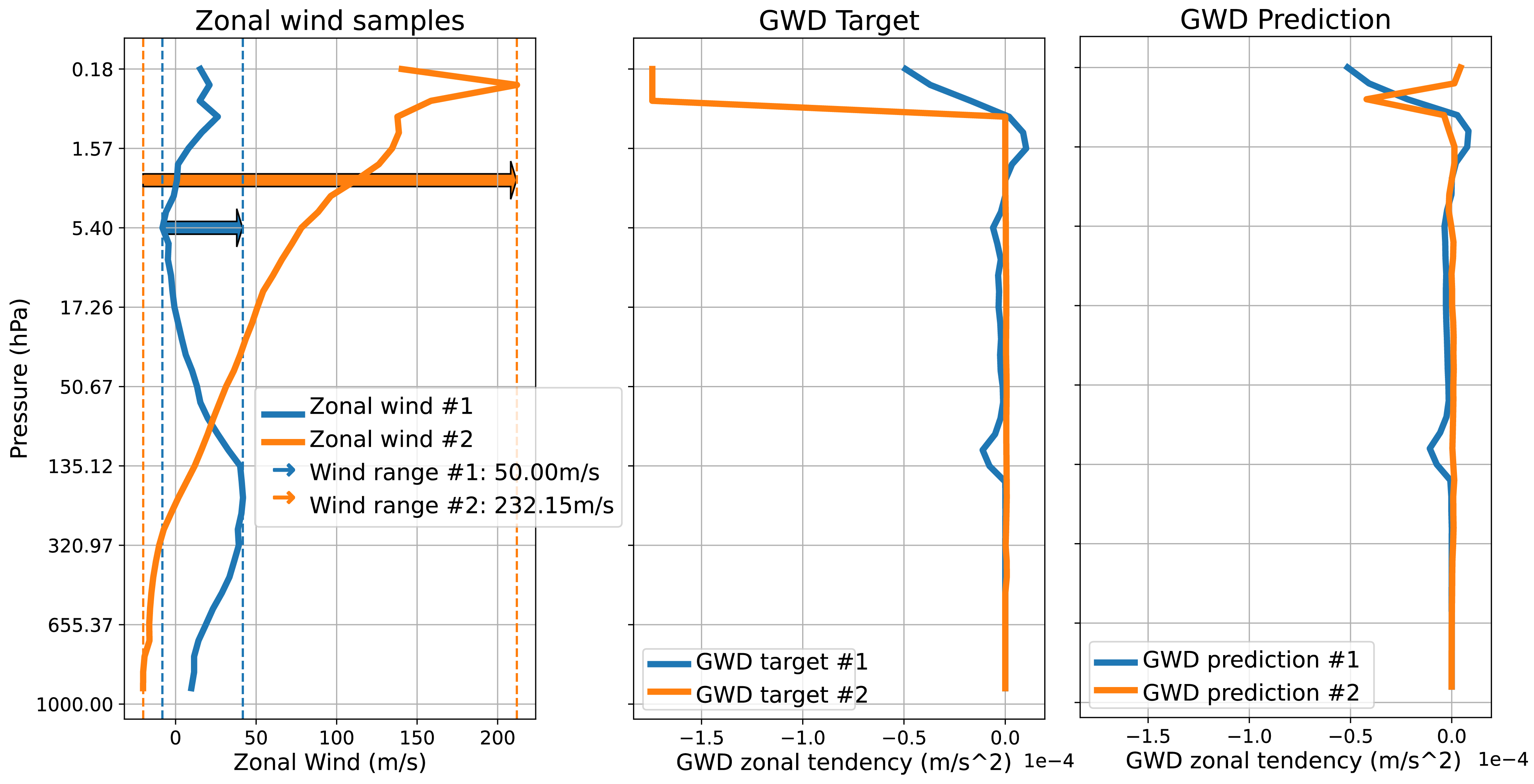}%
  \caption{\label{fig:ex_samples} Left: Two zonal wind profiles sampled near the South Pole at different times in the control integration; Middle: The physics based (AD99) computation of gravity wave momentum deposition (GWD) associated with these two profiles in the left panel; Right: The GWD output by the WaveNet emulator of AD99 for the same input profiles.}
  \label{fig:profiles}
\end{figure}

Physical intuition can be garnered from \Cref{fig:profiles}, which shows two example wind profiles from an integration of the MiMA in the left panel, and the momentum tendency computed by AD99 in the center.  
The scheme also uses the temperature profile (not shown) to determine when convective overturning will lead to GW breaking, but winds are the most important for prediction.  
The blue profile exhibits a more typical case; we will define `typical' precisely below.  
Critical line wave breaking leads to deposition of easterly momentum in easterly shear zones, e.g., near 100 hPa, and conversely westerly momentum in westerly shear zones, e.g., near 1 hPa.
The orange profile demonstrates a less typical case with easterly flow in the troposphere below strong westerly shear throughout the atmospheric column.
Westerly waves are filtered out by easterly winds at the source level (hence no westerly forcing), but the easterly half of the spectrum never experience a critical level.
The scheme thus deposits them all near the model top.

The right panel of \Cref{fig:profiles} provides anecdotal evidence that the WaveNet emulator does a reasonable job of capturing the momentum tendencies from the more typical blue profile case, but fails rather spectacularly with the orange profile. 
As detailed by \citeA{connellyregression}, WaveNet is good at capturing critical level behavior, but struggles to capture non-local effects on the momentum tendencies, both the impact of source level filtering and integrated behavior, where an absence of easterly shear allows waves to reach the top.  

We hypothesize that WaveNet's emulation of AD99 in MiMA suffers from data imbalance, in that gravity wave breaking is most often associated with local critical levels.
WaveNet learns this relationship well.  Cases where the momentum forcing depends on non-local behavior (e.g., when surface level filtering or low level critical levels remove part of the spectrum low in the atmosphere, or when a lack of critical levels leads to momentum deposition near the model top) are more seldom seen, and so tend to be poorly captured the data-driven scheme.
The challenge is to translate this physical intuition into an objective metric to identify the rarer cases dominated by non-local effects.
The input space is 83 dimensional (zonal wind $\vec{u}$ and temperature $\vec{T}$ at 40 levels each, plus surface pressure, latitude, and longitude), but we want a single metric to sort the data. 
After significant trial and error we developed a simple ``wind range" metric that captures many of these rare cases.

\begin{figure}[ht]
  \includegraphics[width=\textwidth]{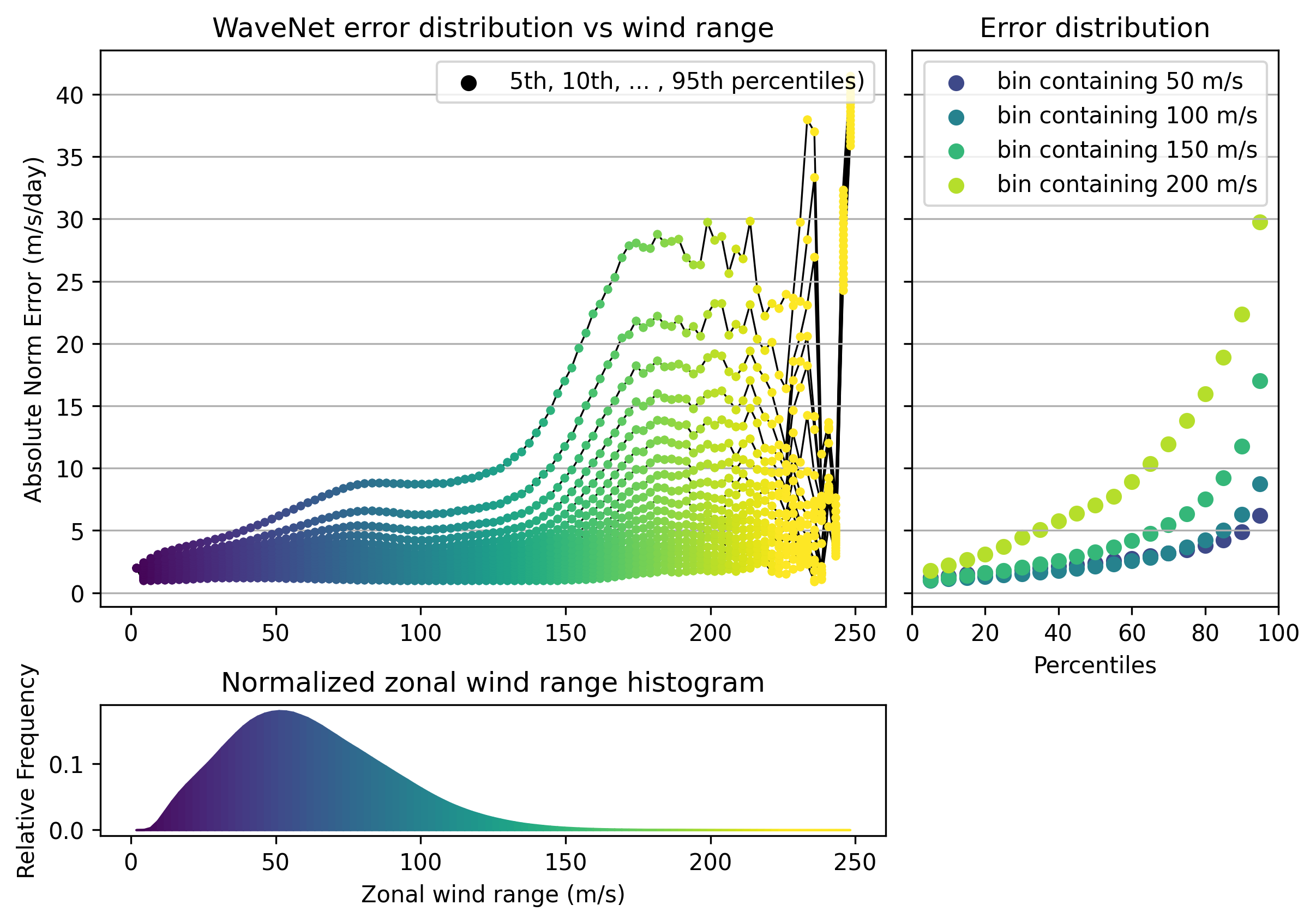}
  \caption{
  \label{fig:imbalanced} 
  Bottom panel shows the histogram of the dataset where each sample is represented by its zonal wind range \cref{eq:windrange}. Frequency is the number of samples in a bin relative to the total number of samples.
  Top left: For each of the 100 equal-width bins of the histogram, we show 5th to 95th absolute error percentiles at 5-percentile increments. Thus we can view the error spread as a function of wind range. Due to noisy error statistics for samples with wind range \textgreater 200 m/s, we exclude those samples in the analysis in the following sections.
  Top right: The error percentiles for a select few bins show that larger errors are incurred more often as the wind range increases.
  }
\end{figure}

The wind shear is a crucial quantity in computing GW forcing on the mean flow. Large shear at any given level favors wave breaking, as GWs over a wider range of phase speeds will experience a critical level.  
Profiles with large shear, particularly at lower levels, tend to exhibit non-local behavior, as the GW spectrum is rapidly depleted, rending upper level critical levels moot.  
(This is to say, a second shear zone will not be associated with GW breaking because waves have already broken below.) 
In addition, strong shear in one direction can lead to cases like that exhibited in Figure \ref{fig:profiles}, where the momentum conservation criterion leads to momentum tendencies near the model top, even if individual waves wouldn't otherwise break there.
An admittedly crude proxy metric we consider to represent the overall presence of shear is the wind range, the total span of winds throughout the atmospheric column.  
Formally, 
\begin{equation}
\mbox{wind range}  = \left(\max_{i=1,\cdots,\tt{nlev}}{u_i}\right) - \left(\min_{i=1,\cdots,\tt{nlev}}{u_i}\right). 
\label{eq:windrange}
\end{equation}
The wind metric is illustrated by the arrows in the left panel of \Cref{fig:profiles}. It  suggests that WaveNet may struggle when the wind range is large (the orange profile).  
While this metric was motivated by the physical argument that these high shear cases are more challenging to learn due to non-local effects,  \Cref{fig:imbalanced} shows that these high wind range cases are rare as well.


The wind range exhibits the two key features of data imbalance. First, the input data exhibits a long tailed distribution with respect to the wind range, and second the ML based emulator systematically struggles with the tail of this distribution. 
This is most clearly illustrated in \Cref{fig:imbalanced}, which shows the distribution of errors for different values of the wind shear.  
The spread of error increases superlinearly with respect to wind range. 
For profiles with a wind spread of 50 m/s, at the mode of the distribution, the error is the prediction of the drag is less than 5 m/s/day for over 90\% of cases. 
For profiles with range of 100 m/s, the error rates are only modestly worse, 85\% of profiles exhibit an error less than 5 m/s/day.  
With a wind range of 150 or 200 m/s, however, only 70 and 30\% of the profiles, respectively, can be predicted with an error of less than 5 m/s.  
Error rates at the 90 percentile are associated with 16 and 28 m/s/day, respectively, a full three to five times worse for cases at the mode of the distribution.

\Cref{fig:imbalanced} motivates  another, even simpler approach of addressing data imbalance: bias removal.  The high absolute error rates for rare profiles with large wind range are in part associated with systematic mean biases in the prediction (not shown).  
In general, a well-trained ML scheme will have no bias in the overall mean, but it can systematically under and over-predict profiles with respect to metrics like the wind range.  
For example, it may trivially under-predict the GW tendencies over the main part of the distribution, but massively over-predict the tendencies at the tail.  
As discussed in Section \ref{ssec:RB} one can remove these biases at the time of inference.

For the remainder of the paper, we use the wind range metric, and the data imbalance it reveals, to improve the training and implementation of WaveNet and a related ML scheme.  
These methods are generic, and ready to apply once a user has identified the metric to quantify the imbalance.  
The better one can sort prediction errors in a high dimensional dataset along a single (or at least a small number of) dimension(s), however, the better one is positioned to use these strategies to improve the scheme.

\section{Treating data imbalance} \label{sec:methods}
Our goal is to help the data driven scheme perform better on the tails of the distribution \emph{without} decreasing performance over the main part of the distribution. 
This makes the typical balancing act between ``bias" and ``variance" that one seeks with any machine learning task more challenging. 
Good performance requires a scheme that both learns the training data well (has low bias) and works equally well on new data (has low variance). 
By this, we mean that the skill is uniform for different samples from the underlying distribution, so it generalizes well to new inputs it has not seen before.  

A large bias is associated with under-fitting, where the method lacks enough training data and/or expressivity to capture the relationships, while a large variance is associated with over-fitting, where the ML scheme uses ``noise" (unimportant features) in the training data to reduce the bias.
This is a case of having too much expressivity relative to the amount of data.
The expressivity of a ML scheme is related to its complexity (roughly, the flexibility it has to identify relationships between inputs and outputs, which is a function of both the method and the number of free parameters it is given).
For our application, we are given some ML scheme of fixed complexity (i.e., WaveNet).
We must ensure there is still enough training data in the center of the distribution to avoid under-fitting it, and not too much emphasis on the tails to cause over-fitting.   

Learning from unbalanced datasets is challenging.
For example, consider a dataset where 99\% of the dataset is class A and the remaining 1\% is class B.
A binary classifier that always predicts class A can still be considered very good under a seemingly innocent metric such as average accuracy, defined as 
\[
\text{average accuracy}\equiv \frac{\# \text{correctly labeled samples}}{\# \text{of total samples}},
\]
with a value of $0.99$, although it completely fails to learn the characteristics of class B. 
Methods to remedy difficulties attributed to imbalanced datasets for classification are far and plenty \cite{he_learning_2009,johnson_survey_2019}, and are used in a variety of applications including object detection \cite{oksuz_imbalance_2021}.

These methods can be broadly categorized into data-level, algorithm-level, and the hybrid of those two.
Data-level methods manipulate the distribution of the training data distribution: such as undersampling from the majority class and oversampling from the minority class \cite{chawla2004special}, or generating synthetic samples of the minority class \cite{chawla2002smote} through randomly weighted linear combinations of samples.
Algorithm-level methods adjust the learning algorithm to increase/decrease the impact of samples from minority/majority class. 
The latter case falls under cost-sensitive learning as it is implemented by imbuing a cost or penalty term in the learning process \cite{krawczyk_learning_2016,elkan2001foundations}.

Although many methods for treating data imbalance are established for classification, extending them for regression is nontrivial. 
There have been some efforts on this front as done by \citeA{Torgo2015,ding_modeling_2019}; and \citeA{rudy_output-weighted_2023}.
\citeA{Torgo2015} extends the Synthetic Minority Oversampling TEchnique (SMOTE; \cite{chawla2002smote}) to regression by assuming near linearity of the model being learned, \citeA{rudy_output-weighted_2023} extends relative entropy based loss functions from scalar outputs to low dimensional vector outputs, and \citeA{ding_modeling_2019} proposes a new loss function and a model design that memorizes extreme events for time series applications. 
Some shortcomings of these solutions are that they are incompatible with nonlinear problems and difficult to implement in applications with high dimensional datasets.

We prepare two methods to address data imbalance in regression tasks.
Both methods require first identifying a metric along which the high-dimensional dataset yields a long-tailed distribution; in our case, the wind range.
We project our high-dimensional dataset to the low-dimensional space identified by the metric. 
\Cref{ssec:histeq} shows how histogram equalization can be applied to transform unbalanced distribution to one more uniform.
This idea is closely related to transportation theory (optimal transport), which is the study of allocation of resources with a constraint of cost appended to the transportation of those resources. 
Since we merely intend to modify the data distribution encountered by the training algorithm, rather than to transform the data itself, we drop the transportation cost constraint.
In \cref{ssec:ourmethod}, we describe the data rebalance method, which extends the ideas of over/undersampling methods to treating data imbalance for regression tasks by applying linear transformations to the probability distribution function (PDF) of the dataset.
Finally, we describe mean bias removal in \cref{ssec:RB}.

\subsection{Histogram equalization} \label{ssec:histeq}
Histogram equalization is an image processing method that adjusts the contrast of an image by changing the shape of the histogram of the intensity values, and is the simplest optimal transport method for 1D data.
The extent to which the shape of the histogram is modified is parameterized by $t\in[0,1]$ where $t=0$ yields the original histogram, and $t=1$ a target histogram.
By equalization, we aim for a target distribution that is uniform, with an equal number of pixels in each intensity bin.

\begin{figure}[h!]
  \includegraphics[width=\textwidth]{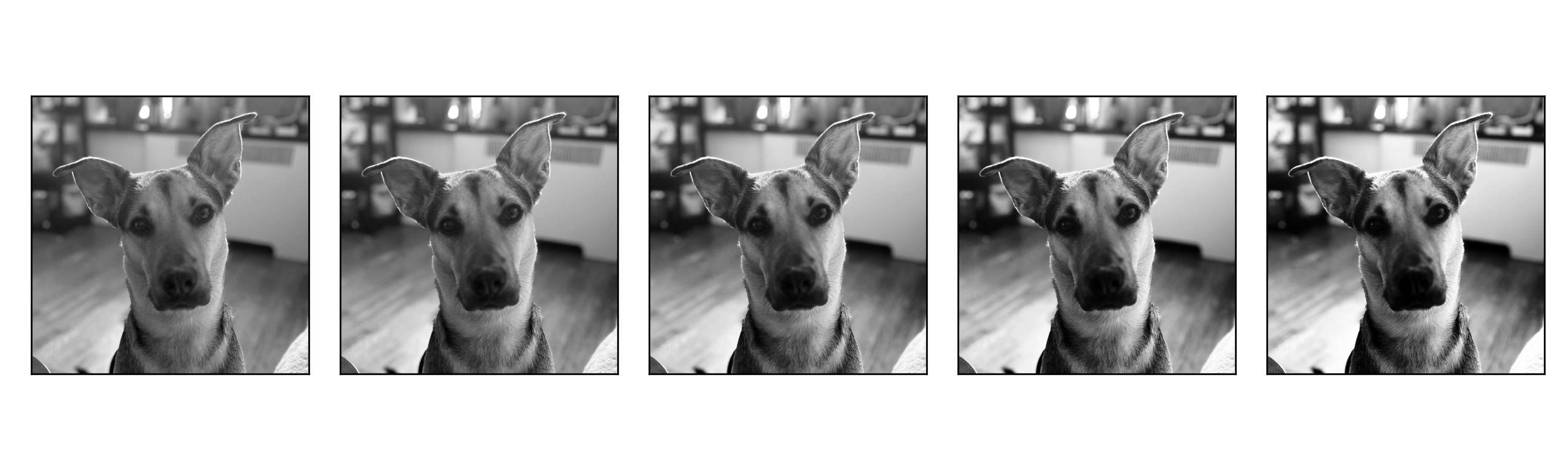}%
  \vspace*{-28pt}
  \includegraphics[width=\textwidth]{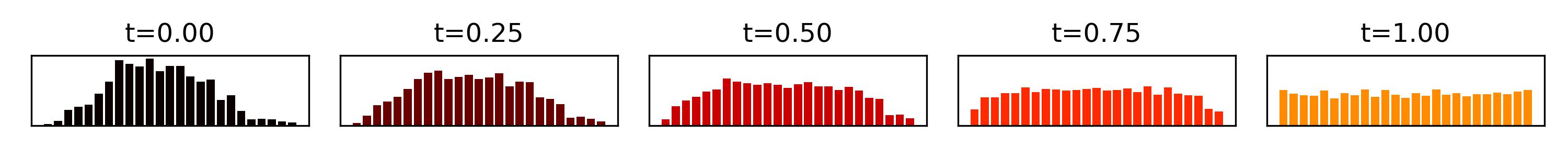}%
  \caption{\label{fig:eq_image} An example of histogram equalization performed for image processing with $t$ ranging from 0 to 1. The original image corresponds to $t$=0. As $t$ increases, moderate saturation pixels are pushed towards their nearest extremes. At $t$=1, the pixels are distributed almost uniformly. }
\end{figure}

\Cref{fig:eq_image} shows an example of this applied to a grayscale image where each sample has a value in $[0,1]$ which represents a greyscale value between black and white. 
The original histogram ($t=0$) has the majority of pixels in the moderate intensity region, and very few pixels are close to minimum and maximum intensities. 
As the parameter $t$ increases to $1$, the distribution is flattened in the peak region and elevated in the extreme regions.
Lighter pixels are made lighter and darker pixels are made darker, qualitatively yielding images with greater contrast as $t$ increases.

Let us describe this procedure in more detail.
Let $x_i$ denote the intensity of the $i$th pixel of an $m\times m$ image, and let permutation $\sigma$ be defined such that $\{x_{\sigma(j)}\}_{j=1}^{m^2}$
 are sorted in increasing order,\[x_{\sigma(1)}\leq\ldots\leq x_{\sigma(m^2)}.\]
Assign $\{y_{j}\}_{j=1}^{m^2}$ to the cumulative distribution function (CDF) of the target distribution.
This corresponds to $m^2$ equispaced, ordered nodes from $0$ to $1$ since the target is the uniform distribution for histogram equalization:\[y_{j} = (j-1)/(m^2-1),\;\; j=1,\cdots,m^2.\]
In general, the CDF of any desired target distribution suffices as the values of $y_j$'s.
Then, the new intensity value for the $i^{th}$ node is given by 
\begin{equation}
  z_i:=(1-t)x_{i}+ty_{\sigma^{-1}(i)}. \label{eq:histeq}
\end{equation}

Here is a numerical example of applying this to a $2\times 2$ image.
The original image is given by pixels
\[\begin{bmatrix}
x_1 & x_2 \\
x_3 & x_4
\end{bmatrix} = \begin{bmatrix}
0.60 & 0.52 \\
0.25 & 0.44
\end{bmatrix}.\]
The sorting permutation is $\sigma=[3,4,2,1]$ for a row-wise uncoiling of the matrix, and the target values are $y_1=0$, $y_2=1/3$, $y_3=2/3$, $y_4=1$. 
Thus, the transformation yields
\[\begin{bmatrix}
y_{\sigma^{-1}(1)=4} & y_{\sigma^{-1}(2)=3} \\
y_{\sigma^{-1}(3)=1} & y_{\sigma^{-1}(4)=2}
\end{bmatrix} = \begin{bmatrix}
1 & 2/3 \\
0 & 1/3
\end{bmatrix}\]
for $t=1$, and the general formula for any $t\in[0,1]$ is  given by 

\[(1-t)\begin{bmatrix}
x_1 & x_2 \\
x_3 & x_4
\end{bmatrix} + t\begin{bmatrix}
y_4 & y_3 \\
y_1 & y_2
\end{bmatrix} =(1-t)\begin{bmatrix}
0.60 & 0.52 \\
0.25 & 0.44
\end{bmatrix} +t\begin{bmatrix}
1 & 2/3 \\
0 & 1/3 
\end{bmatrix}.\]

\subsection{Data Rebalancing}\label{ssec:ourmethod}
Our goal is to change the distribution of the training dataset while taking full use of the available data and without generating synthetic data. 
Histogram equalization for image processing achieves the reshaping of the dataset distribution by transforming the values of the sample from $x_i$ to $z_i$ as shown in \cref{eq:histeq}.
Doing so may move a sample from one histogram bin to another, thereby changing the histogram directly.
Our method uses the linear mapping from the original to the new intensity values described in \cref{eq:histeq}, but apply the mapping to the PDF instead.
The newly assigned probability may increase or decrease a sample's contribution to the training process.
We describe the method in detail below, and propose two implementations of the method in \cref{sssec:weighted_loss,sssec:direct_sampling}, respectively.

Let $H^{(0)}$ be the histogram of the training dataset $X_{\text{training}}$ with $N$ bins,
\[\{[b_0,b_1),\ldots,[b_{N-1},b_N]\}.\]
The count of samples in the $n$th bin, $[b_{n-1},b_n)$ is $h_n^{(0)}$, and the ideal count of the samples in the $n$th bin in the ideal histogram is $h_n^{(1)}$. 
Here, the ideal histogram is uniform with $N$ equal width bins, so $h_n^{(f)}=M/N$ for all $n = 1, \cdots, N$ for a dataset with $M$ samples. 
The new count of the $n$th bin for parameter $t$ is then:
\begin{equation}
h_n^{(t)} =(1-t) h_n^{(0)} + t h_n^{(1)}. \label{eq:newcounts}
\end{equation}
Since the $n$th bin originally represented $h_n^{(0)}/M$ of the training set and now we want it to represent $h_n^{(t)}/M$ of the training set, the ratio between the two determines the resampling rate in the $n$th bin.
\begin{equation}
\alpha_n^{(t)}:=\begin{cases}
h_n^{(t)}/h_n^{(0)}
= (1-t) + th_n^{(f)}/h_n^{(0)}&,\;h_n^{(0)} > 0\\
0&,\;h_n^{(0)} = 0
\end{cases}\label{eq:alpha}
\end{equation}
These ratios determine the new sampling rates for the training data.
We found in practice that fairly low $t$-values still yielded very large $\alpha$ ratios at bins belonging to the extreme tail of the distribution.
To avoid unreasonable resampling rates being assigned to rare data points, we bound the ratios by the maximum repeat parameter as shown in \cref{eq:maxrepeat},
\begin{equation}
    \label{eq:maxrepeat}
    \tilde{\alpha}_n^{(t)}:=\begin{cases}
\min\{\alpha_n^{(t)}, \text{{\tt max\_repeat}}\}&,\;h_n^{(0)} > 0\\
0&,\;h_n^{(0)} = 0.
\end{cases}
\end{equation}

Thus, the final resampling rate, $\tilde{\alpha}_n^t$, is determined by three decisions: 1) choice of histogram bins; 2) $t$, the linear mapping parameter; and 3) the maximum repeat parameter.
The resampling strategy is no longer a simple bilinear interpolation between the original ($h^{(0)}$) and desired ($h^{(1)}$) histograms due the maximum value of the resampling rate.
The counts for the bins of the new, resampled histogram for some $t\in[0,1]$ and {\tt max\_repeat} is,
\begin{equation}
    \tilde{h}_n^{(t)} = \tilde{\alpha}_n^{(t)}h_n^{(0)}.
\end{equation}

\begin{figure}
  \includegraphics[width=\textwidth]{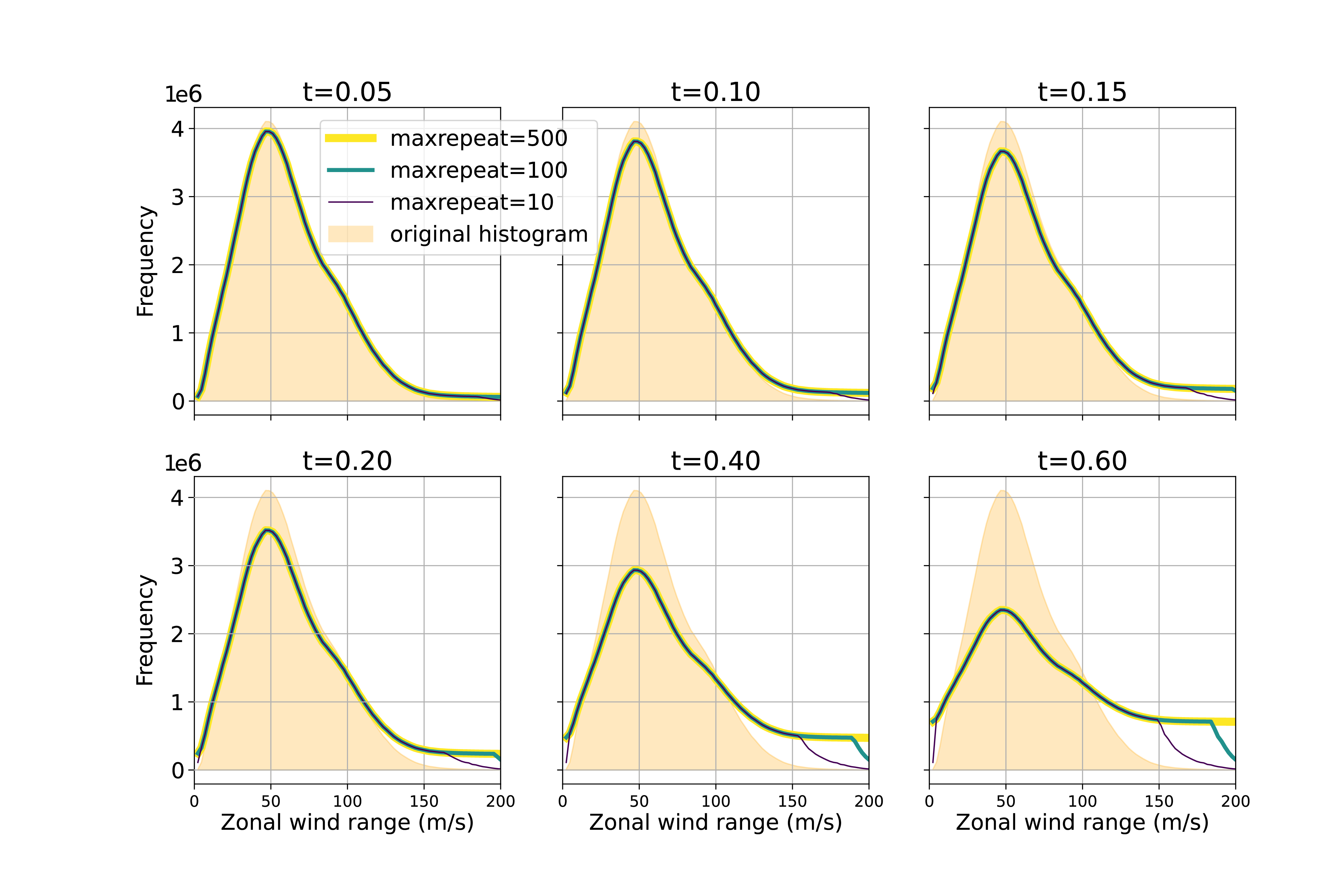}
  \caption{\label{fig:eq_image2}  Each of the panels correspond to $t$ values ranging from 0.05 to 0.60. The 3 lines for each panel represent the impact of maxrepeat parameter values 10, 100, and 500. The original histogram is shown filled in as a basis for comparison.}
\end{figure}

The process is easier to visualize than spell out: \Cref{fig:eq_image2} shows the original histogram, $h^{(0)}$, plotted in foreground with the new histograms, $\tilde{h}^{(t)}$, with increasing values of $t$ for each panel, as well as three different values for {\tt max\_repeat} in each panel.
The impact of {\tt max\_repeat} is seen most clearly in the bottom three panels.
The zonal wind range at which the lower values of {\tt max\_repeat} diverge from the highest value is dependent on $t$ as expected.
The number of histogram bins was kept constant here. It governs how finely one resolves the distribution.
One could also allow the width of the bins to vary, say to more ifnely capture the center vs. the tails. 

\subsubsection{Implementation I: Direct sampling} \label{sssec:direct_sampling}
State-of-the-art optimization methods for deep neural networks rely on incremental, iterative updates of the model weights. They are incremental in that each update is based only a subset of the training dataset called a \emph{batch}, and iterative in that the training dataset is passed through the optimization method many times before the model weights converge to an acceptably optimal state. 
An \emph{epoch} is a measure of unit for the progress of the training of a model defined by a single pass over the training dataset, for which each sample in the training dataset processed exactly once. 
Since our strategy changes the contribution of each sample to the training algorithm based on where in the data distribution the sample belongs, some samples will be seen more often than others.
Therefore, we modify the definition of an epoch to mean a single-pass over a resampled subset of the dataset.
We outline the procedure for resampling in context of a general NN training algorithm, which is written as a pseudoalgorithm (\Cref{algo:training}) in \ref{sec:algos}.

First, compute resampling rates $\tilde{\alpha}_n^{(t)}$ for each bin using \cref{eq:maxrepeat}. 
Next, for each bin labelled by $n=1,\cdots,N$, resample and collect the indices of the chosen samples. 
If $\tilde{\alpha}_n^{(t)}<1$, then it is straightforward to sample from the $n$th bin with probability $\tilde{\alpha}_n^{(t)}$ by randomly choosing a subset of the bin of size $\tilde{h}_n^{(t)}$ without replacement. Another method is to sample from the uniform distribution $h_n^{(0)}$ times and keep the indices that correspond to sampled values less than $\tilde{\alpha}_n^{(t)}$. For both methods, the selected indices are recorded. 
On the other hand, if $\tilde{\alpha}_n^{(t)}>1$, then include every sample from this bin $\mathrm{floor}(\tilde{\alpha}_n^{(t)})$ times, and then sample with probability $\tilde{\alpha}_n^{(t)}-\mathrm{floor}(\tilde{\alpha}_n^{(t)})$.
Following good practice, the collected indices from all $N$ bins should be combined, shuffled, and separated into batches. These batches should then be fed to the training algorithm, which will update the NN model weights once for each batch.

Once all of the batches are processed and if further training is needed, repeat the resampling step to select another realization of the new data distribution.
Note that that every iteration of resampling is done without replacement, but samples may be repeated from one iteration to the next. 
It is straightforward to include an additional step to resample at the next iteration without replacement by keeping track of which samples and how many times those had been picked in previous iterations.
When sampling without replacement is implemented across epochs, all of the samples to be seen by the training algorithm at least once after $\mathrm{ceiling}\left(\left(\min_{n}\tilde{\alpha}_n^{(t)}\right)^{-1}\right)$ epochs. 
We include a pseudoalgorithm for the resampling method in \Cref{algo:sample_bins} in \ref{sec:algos}.

\subsubsection{Implementation II: Weighted Loss Function} \label{sssec:weighted_loss}
An alternative implementation of our approach is to modify the loss function to account for disparity in the distribution.
Success in training deep NNs are attributed to efficient back-propagation, 
a method of updating model weights with the goal of minimizing a loss computed from a batch of samples.
Since loss functions are typically defined for a single pair of the target and the predicted value,
the loss over a batch of samples is an average of the loss function values for each of the samples in that batch.
This implies that every sample in the batch has equal importance in updating the model weights. 
Our resampling strategy aims to modify the data distribution to lend importance to some samples and reduce impact from other samples. 
We propose using a weighted average in the accumulation of loss function values of a batch, where the weight for each sample corresponds to the resampling rate of the bin the sample belongs to. 
For a sample indexed by $i$ that belongs to bin $n$, the weight is determined by parameters $t$ and {\tt max\_repeat} via \cref{eq:maxrepeat}: $w_i \equiv \tilde{\alpha}_n^{(t)}$.
The weights can be computed for the entire training dataset prior to any training and passed to the training loop to compute a weighted average of the loss function for each batch, as shown in \cref{eqn:loss_resampled}. 
\begin{eqnarray}
\mathrm{Loss}_{\text{avg}}(\{y_i\}_{i=1}^{\text{batch size}}, \{\hat{y}_i\}_{i=1}^{\text{batch size}}) &= \frac{1}{\text{batch size}}\sum_{i=1}^{\text{batch size}}\mathrm{Loss}(y_i,\hat{y}_i) \label{eqn:loss_regular} \\ 
\mathrm{Loss}_{\text{weighted avg}}(\{y_i\}_{i=1}^{\text{batch size}}, \{\hat{y}_i\}_{i=1}^{\text{batch size}}) &=  \frac{1}{\text{batch size}}\sum_{i=1}^{\text{batch size}} w_i\mathrm{Loss}(y_i,\hat{y}_i). \label{eqn:loss_resampled}
\end{eqnarray}

\subsubsection{Maximum repeat: Fail-safe against overfitting} \label{sssec:maxrepeat}
The maximum repeat parameter, \cref{eq:maxrepeat}, puts a threshold on the oversampling rate to prevent overfitting.
This allows us to fine tune treatment of the data imbalance
by relaxing the computed resampling rates of bins with high $\alpha$ ratios, which typically occur at the the extreme tail of the distribution. 

\subsection{Bias removal}\label{ssec:RB}
In addition to the resampling method, we propose a correction method to be employed at time of inference to further enhance the quality of the ML model.
This tactic applies a first-order correction to
remedy the bias of a trained model, where the bias is computed along the metric used to identify the data imbalance.
There are a couple of ways to compute the bias.
Consider a dataset of $M$ samples that were binned into $N$ bins where $\mathcal{B}_n$ is the set of indices of samples that belong to the $n$th bin.
The output variable has dimension $d$, and we denote the target and predicted variable of the $i$th sample by
\[\vec{y}_i = \begin{bmatrix}y_{i,1}\\ \vdots \\ y_{i,k}\end{bmatrix},\vec{\hat{y}}_i = \begin{bmatrix}\hat{y}_{i,1}\\ \vdots \\ \hat{y}_{i,k}\end{bmatrix}\]
where $\hat{\cdot}$ is used to denote the ML predictions.
The mean error profile for the entire dataset can be computed by\[
\text{mean error profile}=M^{-1}\sum_{i=1}^M \vec{y}_i - \hat{\vec{y}}_i.
\]
For a well trained scheme, the mean error profile should be close to a vector of zeros.
Similarly, we can compute the mean error profile can be computed for each bin,
\begin{equation}
\text{mean error profile for bin }n
= \left\{|\mathcal{B}_n|^{-1}\sum_{i\in \mathcal{B}_n} \vec{y}_i - \vec{\hat{y}}_i \right\}_{n=1}^N. \label{eq:mean_bias}
\end{equation}
Large errors in bins of the tails can be balanced by smaller errors in the fat pail of the distribution.
At inference, we simply determine the bin the sample belongs to and subtract the appropriate mean bias profile.

\section{Case study: Data-driven GWP emulation}
\label{sec:case_study}
\Cref{ssec:model_arch} describes two model architectures we use to test our method: WaveNet from \citeA{espinosa_machine_2022} and a convolutional NN encoder-dense-decoder (EDD). 
Both implementations of the data rebalancing, with varying $t$ parameters, are applied during training on the same MiMA dataset. 
Offline results are presented in \cref{ssec:offline}, and the emulators with the best offline results are tested online in \cref{ssec:online}.
Here, online refers to replacing AD99 within MiMA integrations with our trained ML emulators. 
\subsection{Model Architectures} \label{ssec:model_arch}
We include a short summary of WaveNet here, and refer readers to \citeA{espinosa_machine_2022} for a full description.
WaveNet takes in a concatenation of all of the input variables and applies several dense layers that split into pressure level-specific ``branches''. 
The branches themselves are also dense layers that output GWD values for a specific pressure level of the MiMA vertical grid, and do not communicate with one another.

The EDD architecture uses 1D convolutional layers in the encoder and decoder sections and dense layers in the middle section.
This structure is imposed to encourage the model to learn local interactions in the encoder section via convolutions while downsampling layers compress the outputs.
This combination of convolutional layers followed by downsampling is commonly used in autoencoders, which can serve as a nonlinear dimension reduction technique that extract essential information. 
The middle dense section allows the processing of global relations and the decoder section reassembles the vertical profile of the zonal gravity wave drag with transposed convolutions and upsampling.
Additional details are included in \ref{sec:arch}.

The hyperparameters for these architectures, listed in \cref{table:model_arch} , include the number and width of the dense layers, the number of (transposed) convolution layers and the size and number of filters for each of these (transposed) convolution layers. 
Some degrees of freedom were removed by restricting the encoder and decoder halves to be as symmetric as possible, while accounting for the fact that the encoder receives multiple channels and the decoder outputs a single channel. 
For the remaining degrees of freedom, we used RayTune (see \citeA{liaw2018tune}) to thoroughly tune the hyperparameters.
We contrast two sizes for each architecture: a smaller network of approximately {\tt350,000} parameters; and a larger network of approximately {\tt700,000} parameters. 
\citeA{espinosa_machine_2022} found that large networks yielded better offline skill than their smaller counterparts, but at the expense of additional computational costs.  
 \begin{table}
 \caption{\label{table:model_arch} Number of trainable parameters in section of each model architecture. The EDD is comprised of 3 sections: encoder, dense, decoder; WaveNet is comprised of 2 sections: shared layers and 33 branches for the top 33 pressure levels. 
 }
 \centering
 \begin{tabular}{l c c c}
 \hline
  Model Type/Size & Convolutional Layers & Dense Layers &  \# Layers per section\\
 \hline
  Small EDD  & 26,237  & 328,800 & 3/3\\
  Large EDD  & 50,337  & 650,800& 3/3\\
 \hline
  Model Type/Size & Shared Layers & Branched Layers &  \# Layers per section\\
 \hline
  Small WaveNet  & 10,368   & 342,177 & 1/3\\
  Large WaveNet  & 14,904  & 704,385 & 1/3
 \end{tabular}
 \end{table}
We present two metrics that are closely related to the mean squared error (MSE), the loss function used during training.

The absolute norm error (AE) is defined as, \[\text{absolute norm error}(y,\hat{y})=\|y-\hat{y}\|_2,\] and was shown, for instance in \cref{fig:imbalanced}.
We also consider the relative norm error (RE) expressed as, \[\text{relative norm error}(y,\hat{y})=\frac{\|y-\hat{y}\|_2}{\|y\|_2}.\]
The relative norm error scales the norm of the error by the magnitude of the target vector, and ensures that the trend of the error norms are not simply proportional to the trend of the target vector norms.
\begin{figure}[ht]
  \centering
  \includegraphics[width=\textwidth]{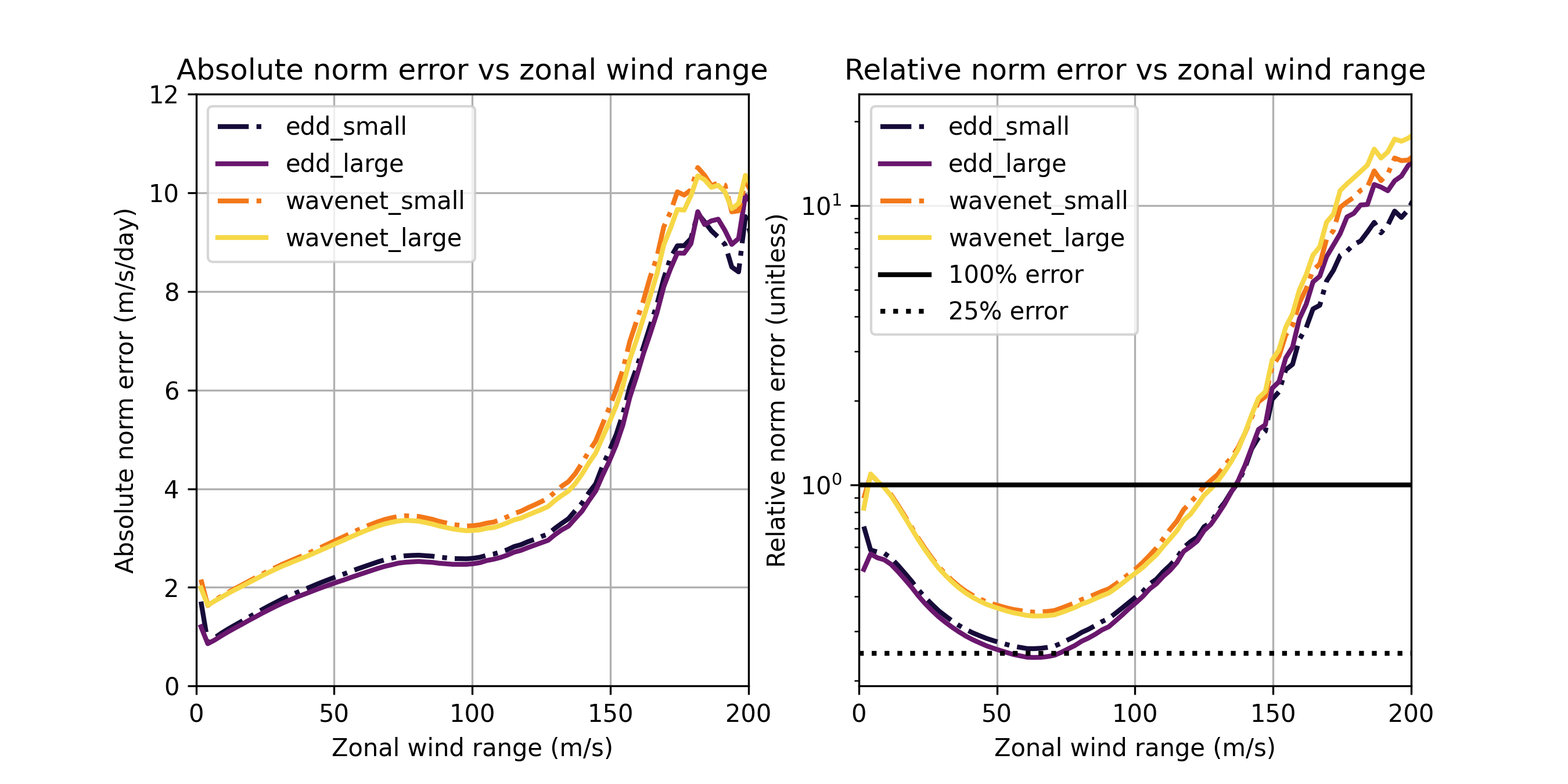}
  \caption{\label{fig:t=0comparisons} Baseline ($t$=0) absolute and relative error norms of two sizes of WaveNet and EDD are shown. 
  The errors are shown as a function of wind range in the validation set, as in \Cref{fig:imbalanced}, which showed results only from the large large WaveNet model. 
  }
\end{figure}

\Cref{fig:t=0comparisons} shows the absolute and relative errors of the four models with no resampling strategy; this establishes a baseline for comparison with our resampling strategies. 
The validation set (data not observed in training) errors are averaged for each bin of the zonal wind range.
We have dropped data points whose zonal wind range are greater than $200$ m/s, as the errors here are too noisy for robust analysis.
We show the relative error on the right panel to highlight how all four variants learn the peak portion of the distribution (refer to the histogram in \cref{fig:imbalanced}) best, but fail at the tail ($>$125m/s).
A relative norm error of 1 (100\% relative error) implies that the magnitude of the error is as large as the target profile itself: a scheme predicting zero drag all the time would satisfy this condition. This suggests that the schemes are doing a pretty awful job for wind range above 125 m/s; predicting nothing at all would be more accurate.

We observe that the EDD models outperform the WaveNet models, albeit the errors are of similar magnitudes.  Overall, the disparity in errors is more significant between the model architectures than between network sizes.
Despite having approximately the same number of learnable parameters as their EDD counterparts, the WaveNet models have not acquired as much skill given identical training conditions; the number of learnable parameters is not all in all when it comes to model complexity. 
The larger variants of both architectures yield smaller average absolute errors than the smaller variants.  The disparity grows slightly for larger zonal wind ranges, though this slight lead of the larger models falters for zonal wind ranges greater than $\approx$ 125 m/s for the relative error.

\subsection{Data Resampling and Offline Results} 
\label{ssec:offline}
Of the three tunable parameters of the resampling strategy, we study the impact of tuning $t$.
The maximum repeat parameter and resolution of the histogram were set at 100 maximum repeats and 100 equal-width bins after an initial survey.
We investigated values of $t=0.05$, $0.10$, $0.15$, $0.20$, $0.40$, $0.60$ following intuition that $t$ closer to $1$ is likely more damaging than helpful given the shape of the distribution of our dataset. 
\Cref{fig:eq_image2} shows the new shape of the data distribution of the 6 configurations on the teal (medium-width) lines with the original distribution shaded in green in the background.

\Cref{fig:large_ae_wl,fig:large_wn,fig:small_ae,fig:small_dnn} show the baseline error ($t=0$, shown in \cref{fig:t=0comparisons}) in black lines, and the deviation of the error relative to this baseline for $t>0$ in colors ranging from brown to yellow.
In all instances we see very little, if any, loss of accuracy in the peak region (a wind range of roughly 10 to 100 m/s).  We have achieved one criterion for success: resampling, either directly or through a weighted loss function, does not damage performance for typical inputs.  Now the harder part: does resampling improve performance in the tail, from 100 to 200 m/s in our wind metric?  Here we found success in most cases, though not uniformly.  We acknowledge the failure first. In our best baseline network, the large EDD, direct oversampling led to overfitting.  In all other cases, however, we were able to successfully reduce error in the tail. 

\subsubsection{Overfitting vs Underfitting} \label{sssec:overfit}
\begin{figure}[ht]
  \centering
  \includegraphics[width=.49\textwidth]{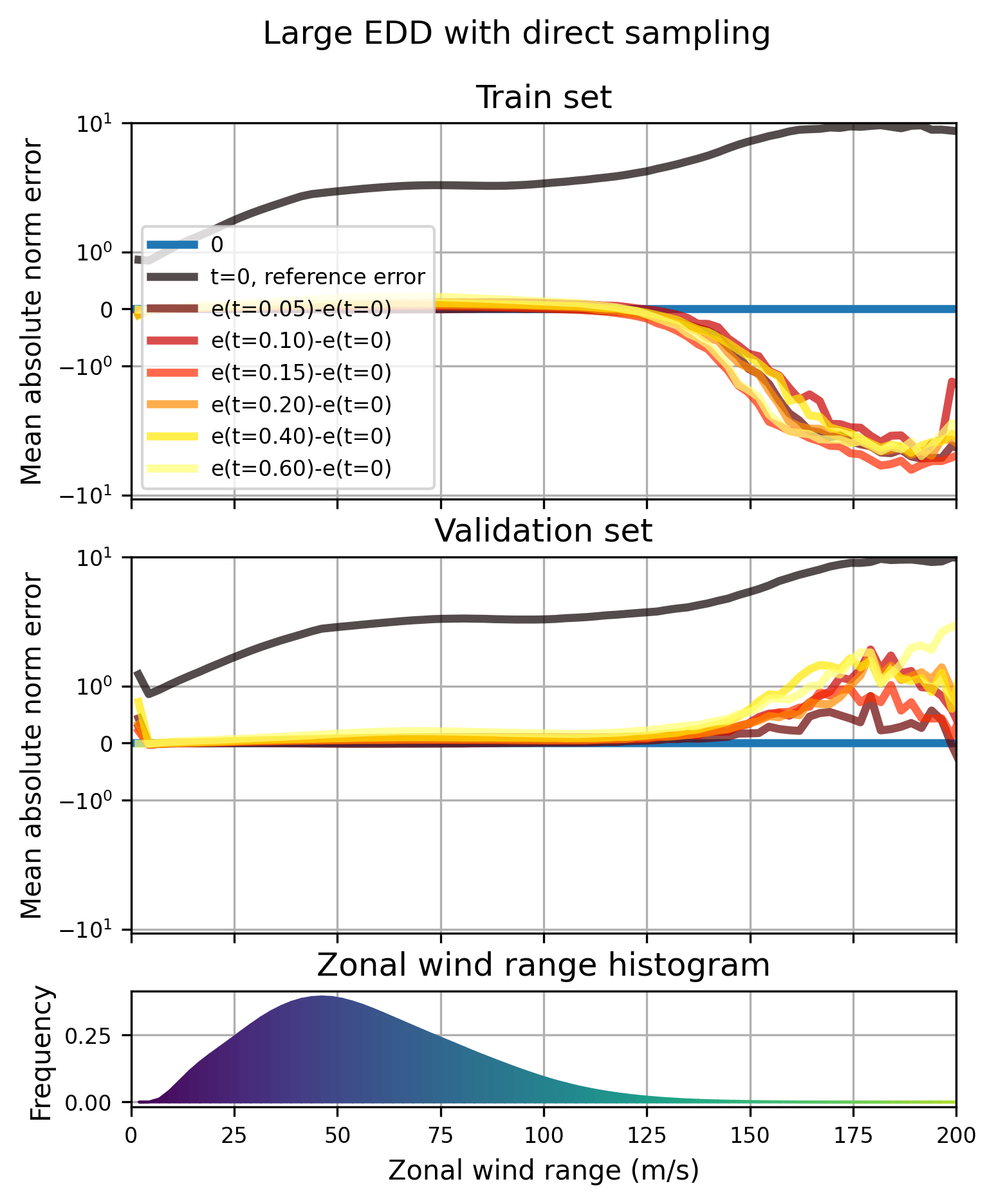} 
  \includegraphics[width=.49\textwidth]{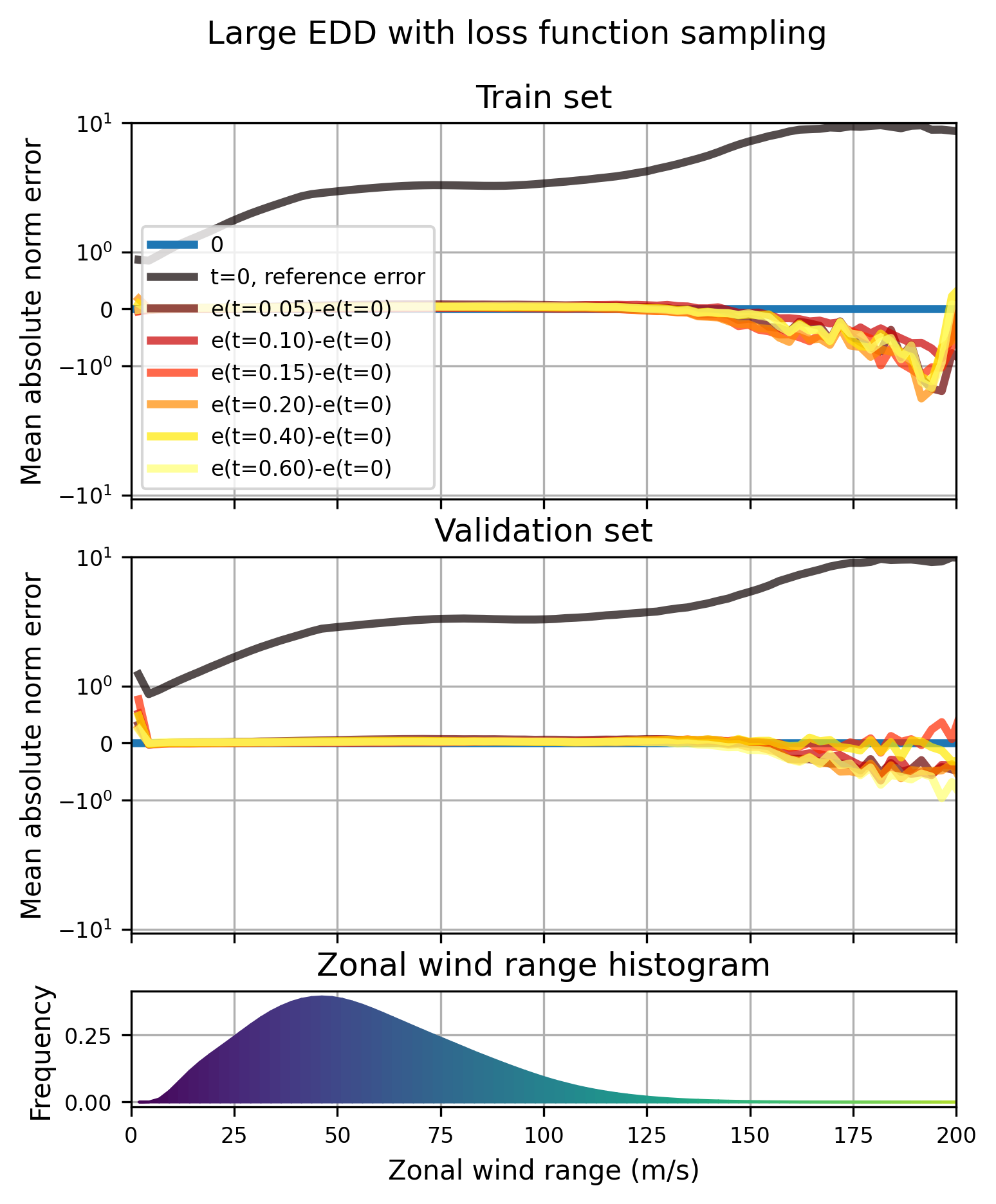}
  \caption{\label{fig:large_ae_wl} This figure show errors from the the baseline ( $t$=0, black ) model and the errors of models trained with the sampling strategy ($t$\textgreater0, colors brighten as $t$ increases) implemented with direct sampling on the large EDD model architecture.
  The top and middle rows show the errors on the training and validation sets, and the bottom row shows the histogram of the dataset with respect to the zonal wind range between 0 and 200 m/s.} 
\end{figure}

As we feared, the resampling strategy can encourage overfitting of the tail in a data driven scheme with sufficient complexity. 
\Cref{fig:large_ae_wl} shows the result of training the large EDD model.
The left panel shows the direct sampling implementation (\Cref{algo:sample_bins}).
For the direct sampling implementation, samples with wind range greater than $125$ m/s \emph{in the training set} suggest impressive gains when compared to the baseline error, albeit with no clear correlation with the $t$ parameter.
This improvement, however, fails to generalize to samples unseen during training: the mean absolute error of the validation set is larger than that of the baseline error.
We observe that larger $t$ corresponds to larger growth in error, suggesting that the trained models suffer from overfitting triggered by the inflation of samples in the moderate tail region.

Typically, overfitting is diagnosed during training when validation error stops improving (or even start to get worse) while training error further improves. 
While we only show the errors at the end of training, it is clear from the design of this experiment that the resampling strategy resulted in models that learned the noise at the tail rather than learning an intrinsic principle tied to the tail. 
A potential cause for overfitting is larger model complexity (number of trainable parameters) relative to the complexity of the pattern being learned, which then leads to the model learning noise associated with the specific instance of the training set.
We suspect that oversampling of the tail combined with the large network size created a learning environment in which the EDD had the capacity to learn noise in the tail.

The right panel of \cref{fig:large_ae_wl} shows the experiment results with the weighted loss implementation (\Cref{algo:weighted_loss}). 
Unlike the direct sampling implementation, we observe about the same magnitude of improvement in the tail for the training set and the validation set.
The upper-middle range $t$-values (0.15, 0.20, 0.40) exhibit no improvements from the baseline in the validation set, but the extreme $t$-values (0.05, 0.10, 0.60) all show slight improvements. 
Since the only difference between the left and the right panels is in the implementation details of the resampling strategy, this suggests that the weighted loss function implementation may be less amenable to overfitting than the direct sampling method.
We further analyze the comparison between the two implementation methods in \cref{sssec:compare_implementations}.

\begin{figure}[ht]
  \centering
   \includegraphics[width=.49\textwidth]{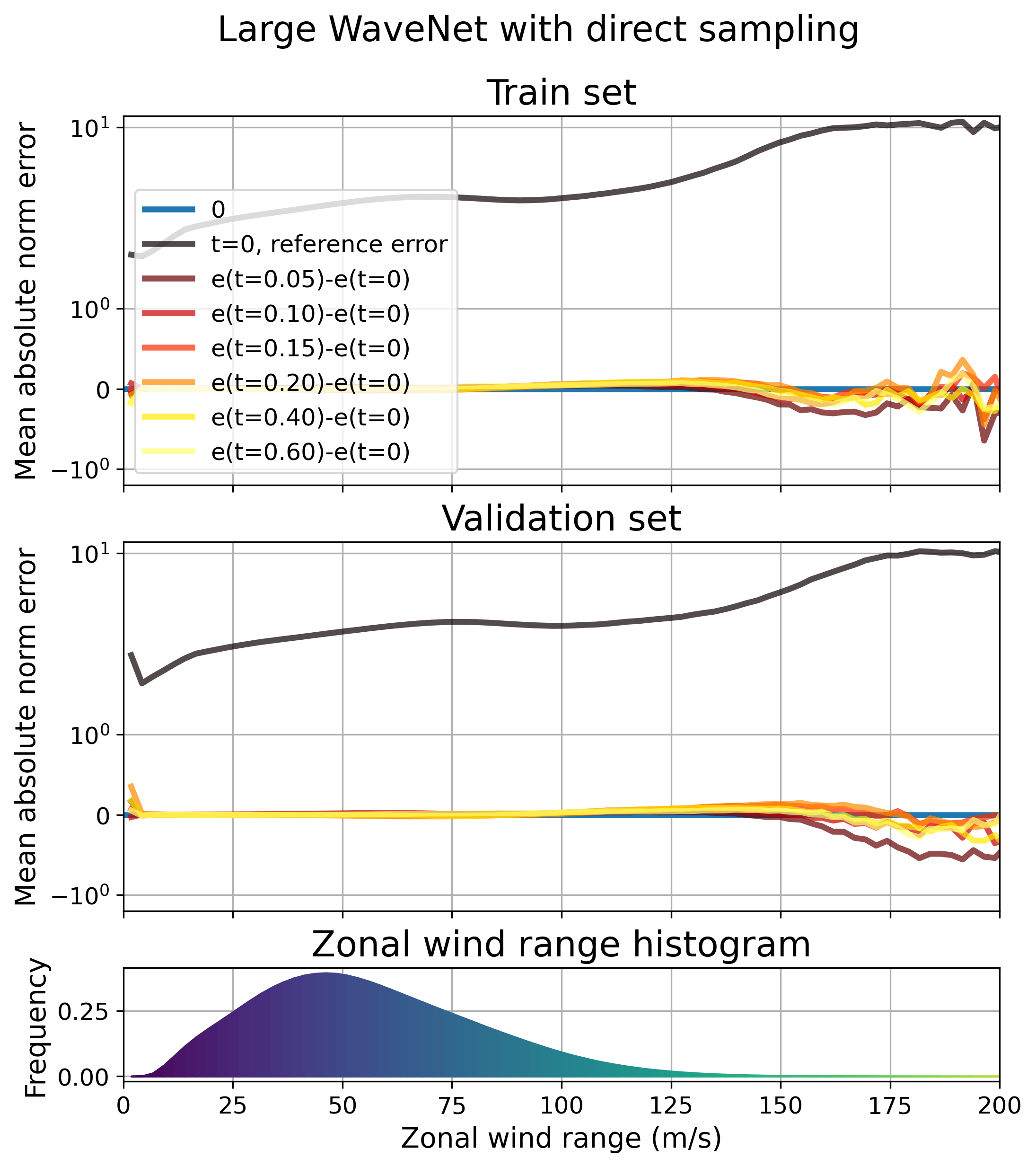}
   \includegraphics[width=.49\textwidth]{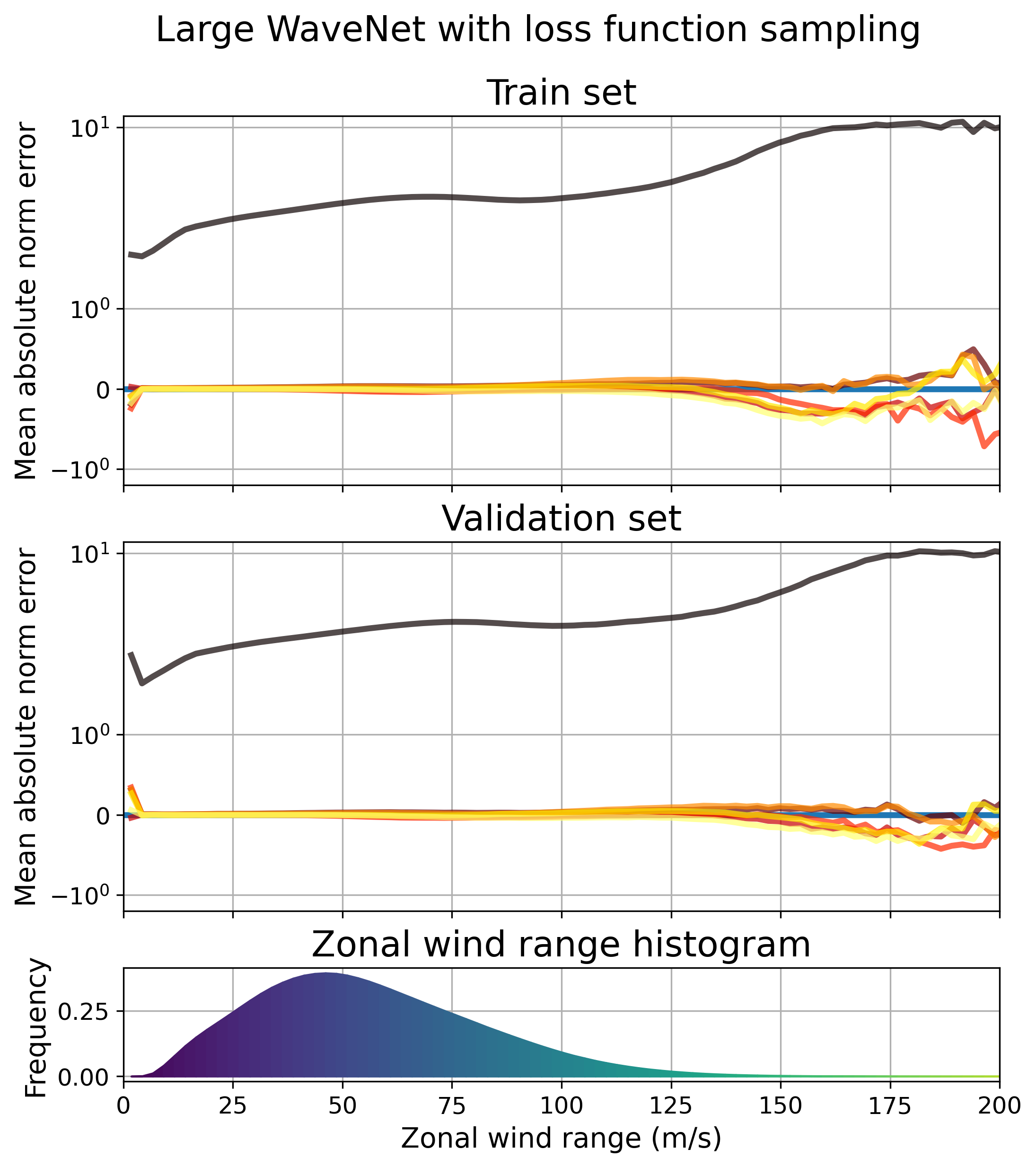}
  \caption{\label{fig:large_wn} Both columns show errors in the same fashion as \cref{fig:large_ae_wl}. Left column shows errors for the large WaveNet instances with direct sampling implementation, and right column shows errors for the large WaveNet instances with weighted loss function sampling implementation.}
\end{figure}

Next, we repeat the experiment in the previous section for the large WaveNet architecture, which has a comparable number of tunable parameters for both implementation methods, and show the result in \cref{fig:large_wn}.
We observe that the validation set errors at the tail are smaller than the baseline error for most $t$ values, and there is no significant change to the errors at the peak.
Unlike the example in \cref{fig:large_ae_wl}, these large networks did not overfit to the samples at the tail of the training set relative to the baseline error. 
If network size is a potential cause for overfitting in the direct sampling large EDD case, why do we not see similar results in the large WaveNet cases? 
We speculate that the baseline WaveNet model was underfitting and there was more room for improvement to be garnered from applying the resampling strategy.
If the baseline EDD model was not underfitting, then the resampling strategy could not reduce the approximation error (bias) much more than was already achieved by the baseline model, and all there was left to learn were noisy traits unique to the training set.

With the exception of the overfitting case, the resampling strategy successfully reduces underfitting at the tail without penalty in the peak, thereby reducing the bias overall.
In the next section, we show further evidence of success of the resampling strategy and compare the two implementation methods.

\subsubsection{Sampling strategy comparison: weighted sampling vs weighted loss} \label{sssec:compare_implementations}
\begin{figure}[ht]
  \centering
   \includegraphics[width=.49\textwidth]{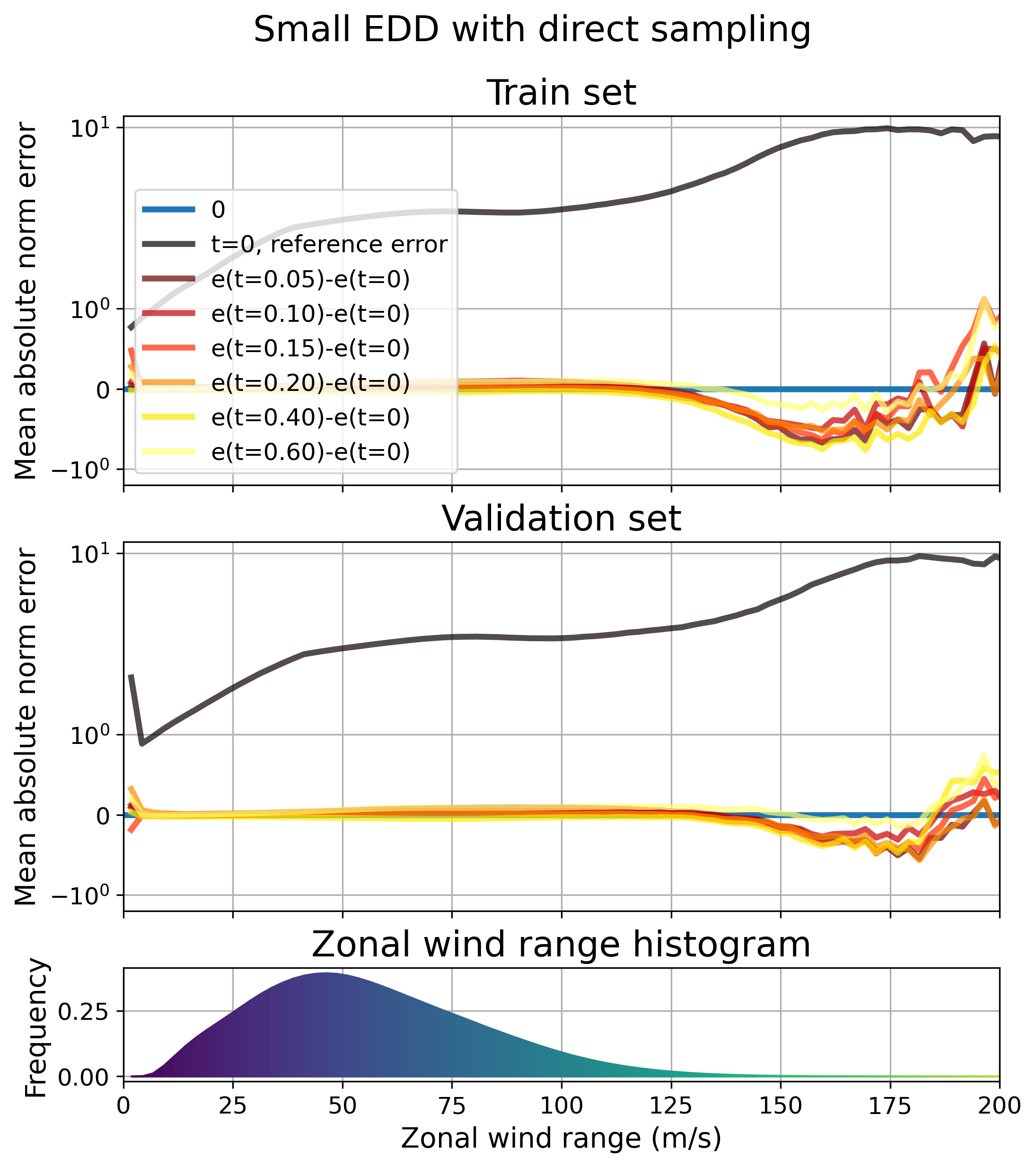}
   \includegraphics[width=.49\textwidth]{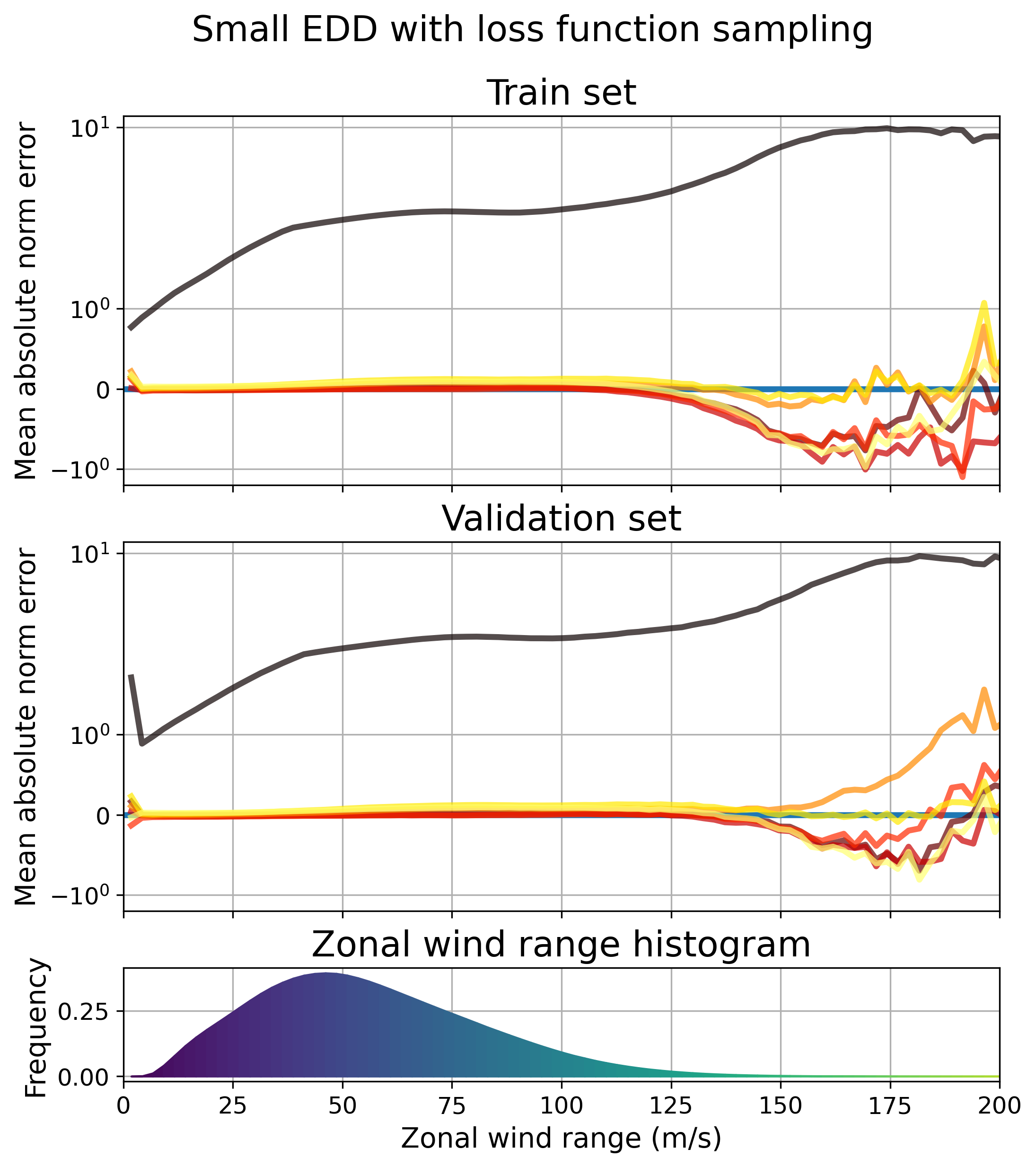}
  \caption{\label{fig:small_ae} Both columns show errors in the same fashion as \cref{fig:large_ae_wl}. Left column shows errors for the small EDD instances with direct sampling implementation, and right column shows errors for the small EDD instances with weighted loss function sampling implementation.}
\end{figure}

We now compare the two implementations (\cref{algo:sample_bins,algo:weighted_loss}) on the small EDD models. 
\Cref{fig:small_ae} shows the baseline errors and the deviations from the baseline errors as we vary $t$ over the training and the validation sets.
\Cref{fig:small_ae} reveals improvements in the tail, albeit modest, with little to no damage in the peak. 
The notable exceptions occur at $t=0.20$ and $t=0.40$ for the weighted loss implementation, where there are almost no change if not a decline in performance on the tail. 
These occur in both the training and validation set, however, and therefore are not likely an issue of overfitting.
Outside of those exceptions, improvements occur for a wider range of the distribution, with larger magnitudes of improvement in the training set than in the validation set as expected. 
The weighted loss experiment (right plots of \cref{fig:small_ae}) shows a slightly larger disparity between the training and validation set errors than the direct sampling experiment; the training set errors show larger improvements with the weighted loss implementation than direct sampling, but the validation errors are comparable between the two implementations.
With direct sampling, all $t$ values except for $t=0.60$ still yield improvement in error in the moderate tail region.

\begin{figure}[ht]
  \centering
  \includegraphics[width=.49\textwidth]{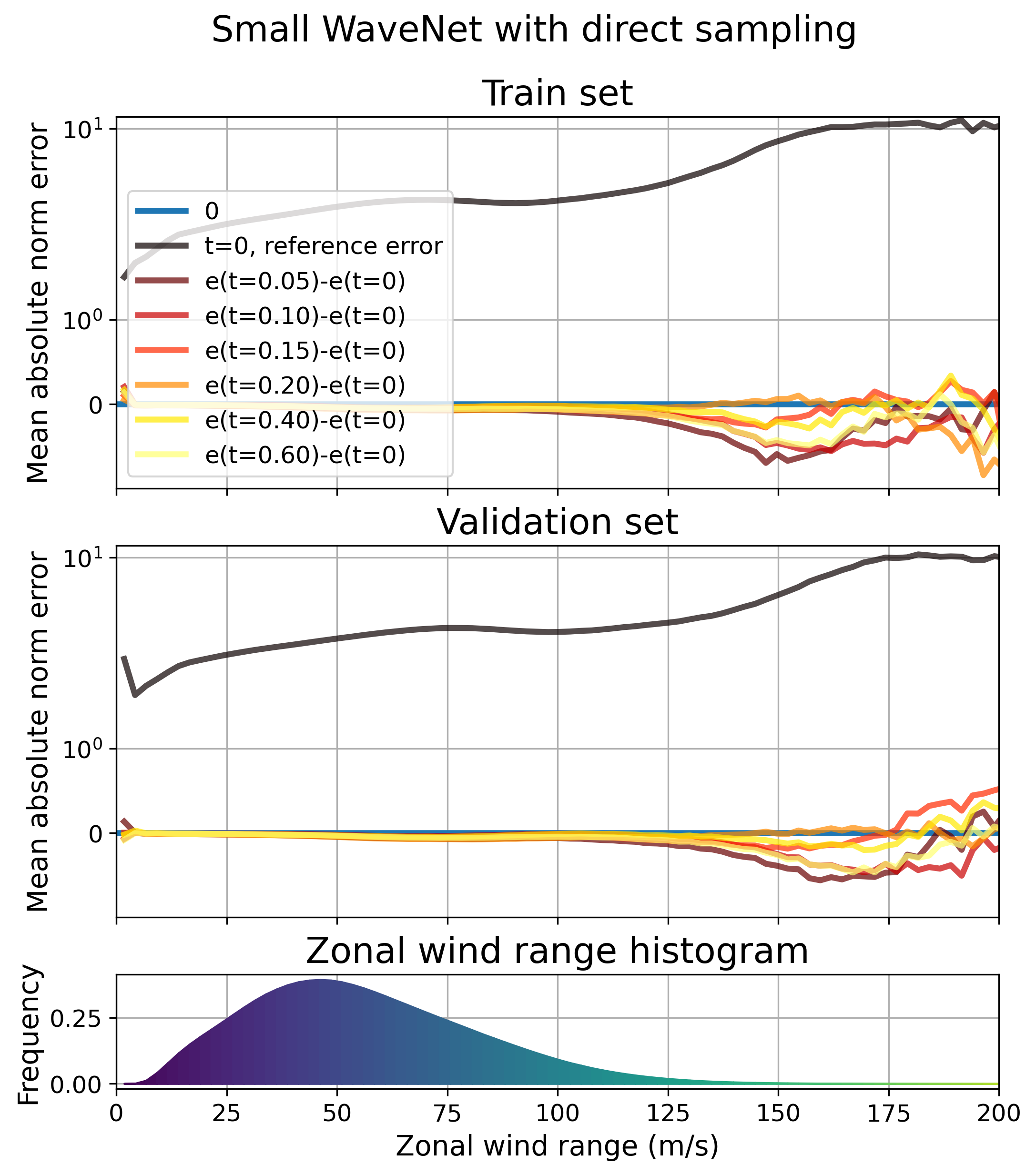}
  \includegraphics[width=.49\textwidth]{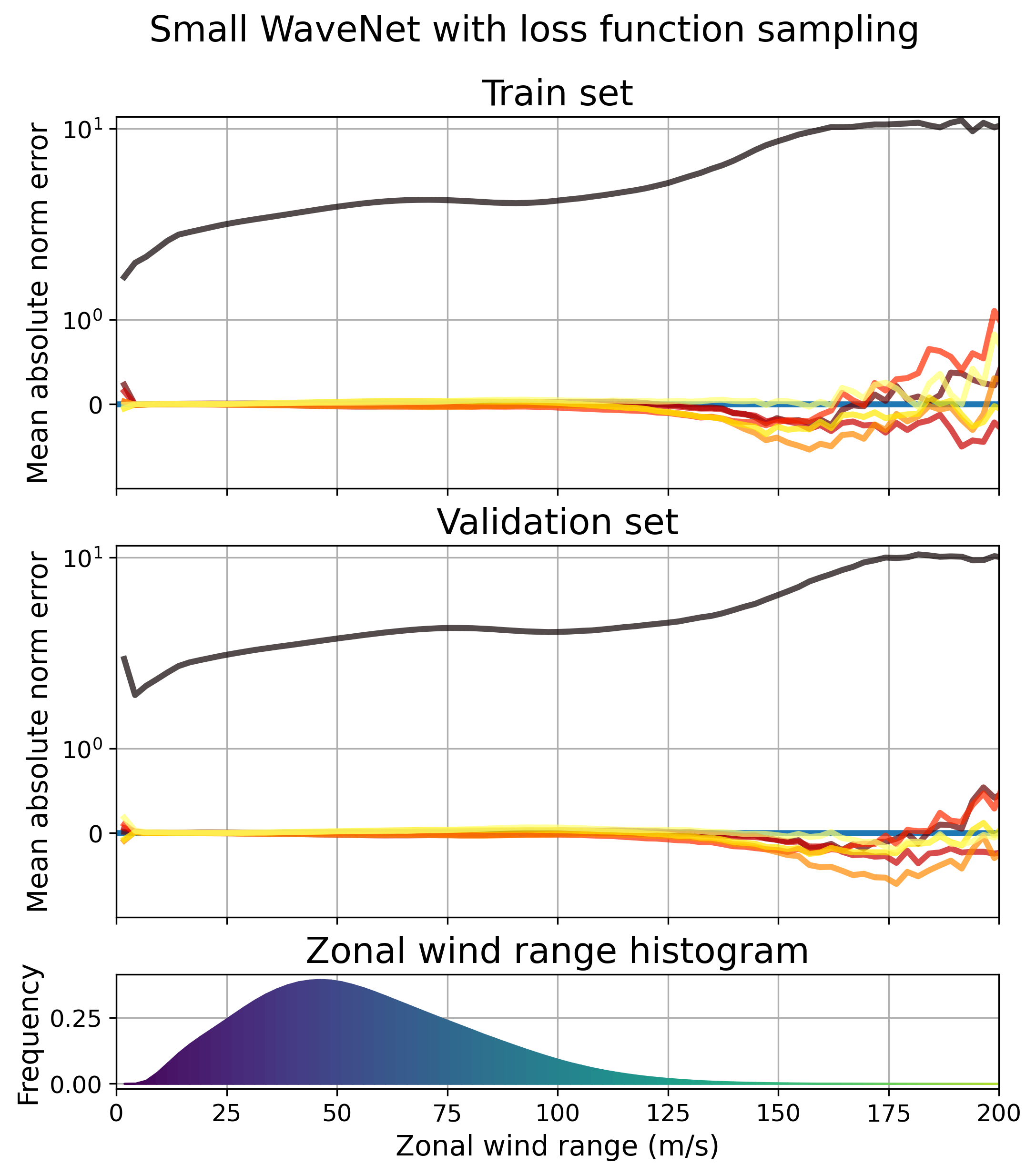}

  \caption{\label{fig:small_dnn} Both columns show errors in the same fashion as \cref{fig:large_ae_wl,fig:large_wn,fig:small_ae}. Left column shows errors for the small WaveNet instances with direct sampling implementation, and right column shows errors for the small WaveNet instances with weighted loss function sampling implementation.}
\end{figure}

Next, we discuss the experiment results for the small WaveNet model. 
As shown in \cref{fig:small_dnn}, the difference between direct sampling and weighted loss are less pronounced than in the EDD model.
Also, the errors of the training set and the validation set are much closer than in the experiments for the small EDD models. 
The largest difference between the implementation methods for the small WaveNet models is in which $t$ values are the most optimal.
The direct sampling method is optimized for the smallest and largest $t$ values, whereas the weighted loss method prefers moderate $t$ values ($t\approx 0.15$).

Even though we saw that the loss function sampling avoided overfitting for the large EDD experiment, we do not see a similar advantage of the loss function implementation over the direct sampling implementation in the small EDD, small WaveNet, and large WaveNet experiments.
However, we do see modest improvements in the tail for models trained with the resampling strategy for the majority of $t$ values for those three experiments, although there is no clear trend of which $t$ values are optimal.
Future experiments that may reveal tighter trends, include studying the sensitivity of learning algorithm, and increasing the density of $t$ values.

\subsection{Bias Removal and Online Results}\label{ssec:online}
We conclude our case study with a brief discussion of how our modified data-driven parameterizations perform when coupled ``online'' with the MiMA atmospheric model.  
An important evaluation of a new parameterization scheme is conducted by computing statistics from long-time integrations where the scheme is coupled with the model, as opposed to the ``offline" metrics we showed in the previous section.  Online coupling is a more challenging task, as errors in the GWP can lead to biases in the large scale flow, forcing the scheme to make inferences in regimes it has not yet seen, which often leads to instability \cite{brenowitz2020interpreting}.

To test a selection of our trained ML emulators, we follow \citeA{espinosa_machine_2022}, coupling them with MiMA for 40-year integrations after 20 years of model spin-up. The simulations with the data-driven emulators can then be compared against the control integration with the ``true" gravity wave forcing provided by the AD99 physics based parameterization.  Coupling also allowed us to implement the bias correction, which can be implemented independently or in addition to the rebalancing strategies.  To summarize quickly, the new data driven parameterizations successfully couple with the model, producing climatological statistics (mean and variability) that were consistent with the original model.  Differences between the model with the baseline schemes and our re-balanced versions, however were not statistically significant.  
It is likely that a longer integration could eventually reveal significant differences, but an improvement that requires a century or more to observe is of modest utility.  We conclude that while re-balancing the data did improve performance based on the wind metric, this bias was either not critical to performance of the parameterization in the model, or we have not sufficiently improved the tails to see a significant effect.  

\begin{figure}[ht]
  \centering
  \includegraphics[width=\textwidth]{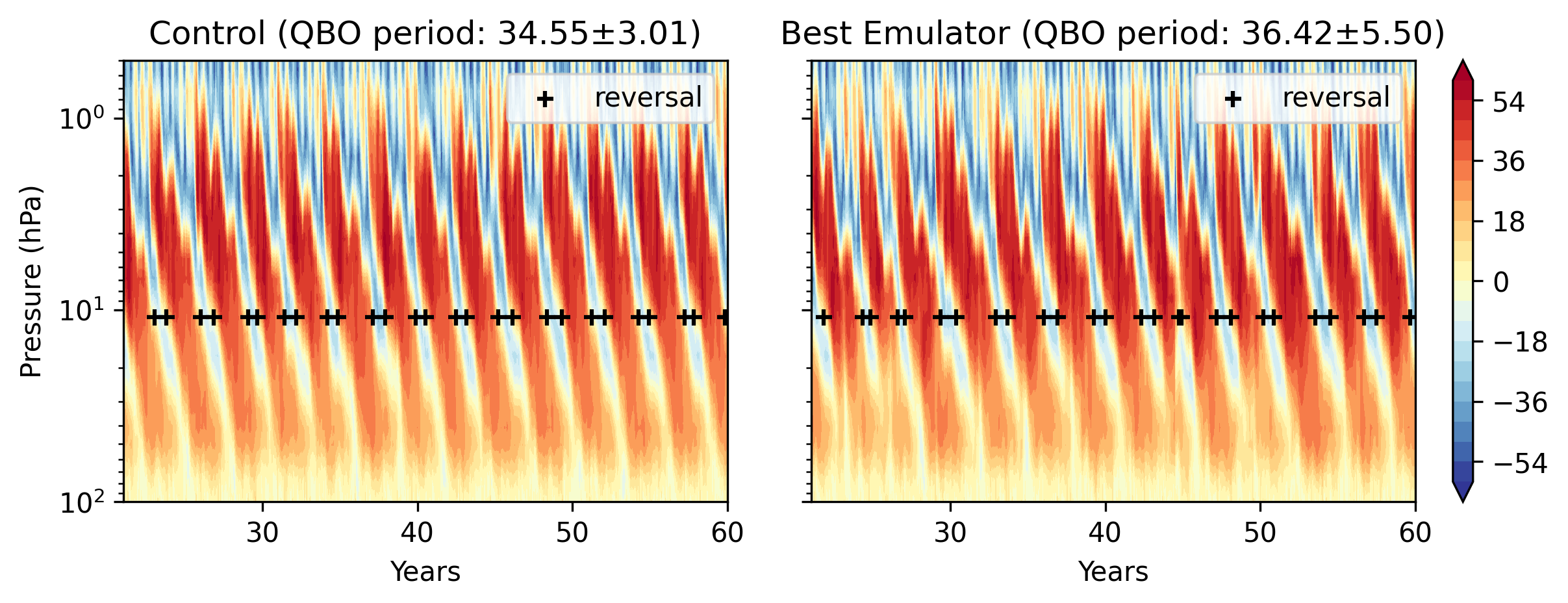}
  \caption{\label{fig:QBO_best} Both plots show the zonal mean zonal wind averaged over years 20-40 in latitudes between $5^{\circ}$S and $5^{\circ}$N. The crosses indicate the times where QBO phase changes are detected by the TT method. Left: Control run with AD99; Right: Best emulator (small EDD with resampling strategy via weighted loss and $t=0.05$ and bias removal). }
\end{figure}

For completeness, we show a few results here, focusing on the coupled model's ability to generate the Quasi-Biennial Oscillation (QBO), a vacillation of easterly and westerly jets in the tropical stratosphere over a period of approximately 28 months.  We highlight this metric because the QBO is in large part driven by gravity wave momentum transport.  This emergent behavior on a time scale of years, generated from gravity waves that operate on time scales of hours, is viewed as critical test of gravity wave parameterizations \cite{richter2022response,anstey2022impacts,bushell2022evaluation}.
An important difference between the online runs in this manuscript and that of \cite{espinosa_machine_2022} is in the model parameters of MiMA that generated the training data. 
We employed parameters that were optimized for simulation of the Northern hemisphere \cite{garfinkel2020building}, not the QBO.  Thus the oscillation is the control integration had a period of approximately 35 months, not 28 months, as shown in \Cref{fig:QBO_best}.  Capturing the right period of the QBO is generally achieved by tuning the GWP, as was done in \citeA{garfinkel2022qbo}.

We show results with the smaller EDD models, as the rebalancing strategies exhibited the largest offline improvement. \Cref{table:QBOs} lists the QBO period for the baseline model ($t=0$) and the various combination of resampling strategy and bias removal.
The QBO period 
was computed using the Transition Time (TT) method of \citeA{richter2020progress}. 
First, the zonal wind was averaged zonally in the tropical region (latitudes between $5^{\circ}$S and $5^{\circ}$N), as shown in \Cref{fig:QBO_best}.
Then the intervals between QBO phase changes are defined as times when the signs of zonal mean zonal wind reversal near 10 hPa (denoted by the plus signs).
The resulting mean and the standard error of those values give us a proxy for a confidence interval of the QBO period.
A robust implementation of the TT method requires smoothing the field with $15$ to $30$ day windows, to avoid double counting small deviations around transitions.

 \begin{table}
 \caption{\label{table:QBOs} }
 \centering
 \begin{tabular}{l c c c}
 \hline
 Emulator Description & Transition Time \\
 \hline
 control & $35.01\pm2.46$\\
 small EDD, $t=0$ & $38.37\pm6.59$\\
 small EDD, $t=0$, bias removed & $37.01\pm7.70$\\
 small EDD, $t=0.05$, direct sampling & $37.05\pm2.98$\\
 small EDD, $t=0.05$, direct sampling, bias removed & $38.12\pm3.69$\\
 small EDD, $t=0.05$, weighted loss& $39.66\pm8.97$\\
 small EDD, $t=0.05$, weighted loss, bias removed& $36.42\pm5.47$\\
 \end{tabular}
 \end{table}
The baseline model exhibited a slightly longer QBO period of 38 months, though 40 years of simulation was insufficient to establish whether this bias is statistically significant.  We found that all of our modified data-driven approaches exhibited shorter QBO periods, an improvement relative to the baseline, but still biased long relative to the control.  The best performing model is highlighted in \Cref{fig:QBO_best}, but as quantified in \Cref{table:QBOs}, these integrations are not long enough to establish whether these differences are statistically significant.  As noted above, this could be due to the fact that the QBO bias is unrelated to errors in the rare cases highlighted by the wind metric, or that our correction is insufficiently large to make a dent.  It highlights the importance of domain knowledge to identify the key metric(s) of data imbalance that matter for the problem of interest.

\section{Conclusions and Future Directions} \label{sec:conclusion}
With the growing prevalence data-driven methods being used for various tasks in modeling earth system models, it is crucial to properly learn from geoscience datasets.
We address what one can do to improve a data-driven parameterization given that there is no additional data to learn from, nor computational capacity to allow for a larger, more complex model. In other words, this is the typical scenario for modeling various subgrid-scale mechanisms in climate models. 
In particular, we proposed two strategies to combat data imbalance with the goal of improving data-driven models, and applied it to a case study of improving a data-driven GWP model.

Both methods rely on first identifying a metric or a projection that yields reveal an imbalance in the available dataset that has an inherent significance to the physical process being modelled. 
This process is unique to each application and requires expert scientific knowledge of the modelled process, and doubles as a dimension reduction step that allows the practitioners to view the original high-dimensional dataset in a new context.
Ideally, this new context should illuminate the differences between frequent (and therefore easy to model) instances from rare (and difficult to model) instances.
Despite resulting from the same physical mechanisms, these two types of instances occupy almost two distinct regimes due to the natural variability in the model system.
A necessary complicating factor is that these two \emph{classes} are not sharply partitioned like discrete distributions, but rather can be viewed as the peak and the tail of a continuous distribution.
In our case study, we chose wind range of a model column as the appropriate metric for our physical process, gravity waves.
This choice stemmed from the observation that wind range can crudlely approximate shear, an important quantity in determining the level at which GWs break.


Data rebalancing can be achieved in two ways. In the first method, we use the distribution of the dataset along the identified metric to systematically undersample from the peak and oversample from the tail. 
Our motivation to undersample from the peak is from the intuition that these samples are over-represented relative to the variability they cover within the dataset, resulting in trained models that may overfit to this region.
On the other hand, oversampling the tail is justified by the exact inverse logic: these rare samples are undervalued in their influence over training models.  In the second method, the sampling is left unchanged, but the loss function is weighted by the same ratio to increase the penalty on the under-represented class and reduce it for over-represented class.
Both data rebalancing strategies generate a new distribution/weighting function on the training dataset with a linear interpolation of the original distribution to a desired distribution (i.e., uniform distribution) parameterized by $t\in[0,1]$, much like histogram equalization.
We add in an additional parameter to prevent too much oversampling/weighting in the tail, by the name of maximum repeat. 
The implementation of these methods requires discretizing the continuous distribution into discrete bins; the choice of the histogram is also an important choice.
The methods are implemented by either providing to the learning algorithm a subsample of the dataset that realizes the new distribution, or using the new weights in the loss function such that the new distribution is implicitly represented.

In our case study, we found that data rebalancing successfully reduces the errors in the moderate tail region while maintaining approximately the same error levels in the peak under most scenarios.
In the exception case, data resampling increased the generalization error in the tail, which we attribute to the large size of the ML model.
Too large of a model complexity can cause a model to learn noise rather than pattern in the dataset, a phenomenon exacerbated by oversampling in the tail. 
Unfortunately, we do not observe a clear advantage of the direct sampling implementation over the weighted loss implementation, nor an unambiguous indication of how to choose the method parameters. 
Further studies are needed to address these issues on well-understood datasets: the dataset used for our case study is likely not the best tool for developing intuition for this method.

Mean bias removal, an additional approach to fix errors with data imbalance, corrects the extant bias in a fully trained data-driven model as a function of the data imbalance-revealing metric
This is a first-order correction as it assumes that the mean bias profile of the trained model evolves meaningfully across this metric. 
The main source of error for this method is generalization error as the mean bias profiles of the training set may not be representative of the instances available at time of inference.

In conclusion, data rebalancing and bias removal show modest improvements in producing data-driven models less inclined to mirror the imbalance apparent in the dataset.
The lack of overwhelming evidence of the success of these methods can be attributed to several factors. 
First, our research did not investigate how to choose the projection used to identify data imbalance, a crucial component to both data rebalancing and mean bias removal. 
Thus, it may be that the wind range metric is not the most ideal projection for the dataset used in our case study, or that any 1D projection is too simple to capture the data imbalance for this dataset.
Second, our assumptions on how the data imbalance impacts the training of the data-driven models may be overly simplistic, especially in its treatment of the tail. 
We view samples from the tail as in need of a greater significance in training the ML model. 
However, a more pressing issue at the tail may be that the dataset available to us does not cover the variability inherent to that region.
If so, any oversampling does not increase coverage in this region but instead lead to overfitting. 
We attempt to curb this by introducing the maximum repeat parameter, but this introduces another parameter to be tuned in the rebalancing method. 
Scarcity of rare (and extreme) phenomena in datasets is a common challenge in geoscience datasets that may be alleviated by rare event sampling, and beyond the scope of the methods presented in this paper.
\section{Open Research}
\subsection{Data Availability}
All neural networks used in this manuscript were (re-)written in PyTorch \cite{paszke2019pytorch}. 
The WaveNet implementation in PyTorch exactly followed the descriptions in \cite{espinosa_machine_2022}.
Model of an Idealized Moist Atmosphere (MiMA) \cite{Jucker2017,garfinkel2020building} is maintained at https:// github.com/mjucker/MiMA and available at https://doi.org/10.5281/zenodo.3984605.
The model code, forpy coupling code, trained ANNs, run parameters, and modified configuration for MiMA are available at \url{https://github.com/yangminah/GWPRebalance}.  
The coupling library, forpy, developed and maintained by Elias Rabel is well documented and available at \url{https://github.com/ylikx/forpy}.


%


\acknowledgments
This work was supported by the U.S. National Science Foundation through award OAC-2004572 and Schmidt Sciences, as part of the Virtual Earth System Research Institute (VESRI).
The manuscript benefited greatly from conversations with Joan Alexander and Pedram Hassanzadeh.  We also thank the NYU High Performance Computing center, where the model integrations were performed.    

\clearpage


%
\bibliography{new,zotero_lib} 




%
%
%
%
%
\appendix
\section{Formal Algorithm Details} \label{sec:algos}
\Cref{algo:training} shows an example of how to incorporate the direct sampling implementation of the resampling strategy within the framework of any stochastic gradient descent-type learning algorithm that processes batches of training samples at a time. 
Next, \Cref{algo:sample_bins,algo:weighted_loss} show the direct sampling and weighted loss sampling implementations in detail. 
\Cref{algo:training} can easily be modified to use \Cref{algo:weighted_loss}, where the computed weights are passed into the loss function in the optimization step in line 6, and lines 1, 3, and 4 can be omitted. 
\begin{algorithm2e}[h]
  \DontPrintSemicolon
  \KwIn{$\X$, Training set; $\hat{\varphi}$, machine learning model; $\{C_n^{(1)}\}_{n=1}^N$, counts of bins of ideal histogram; $t$, linear parameter; {\tt max\_repeat}, maximum repeat parameter.r.}
  $\{I_n^{(0)}\}_{n=1}^N \gets $ Bin $\X$ into $N$ bins. \tcp*{$I_n^{(0)}$ is the list of indices in the $n$th bin. }
  \While{$\hat{\varphi}$ needs further improvement}{
  \tcp{This while-block encompasses a pass over the training set.}
    $I^{(t)}  \gets \tt{resample(}\{I_n^{(0)}\}_{n=1}^N, \{C_n^{(0)}\}_{n=1}^N, t, \tt{max\_repeat)}$\\
    Shuffle $I^{(t)}$ and divide it into $B$ batches ($I^{(t)}=\cup_{b=1}^B I_b$). \\
    \For{b=1:B}{
      Optimize $\hat{\varphi}$ over $\X[I_b]$.
      }
    }
  \Return $\hat{\varphi} $ \tcp*{Trained model}
  \caption{Training structure.}
  \label{algo:training}
\end{algorithm2e}

\begin{algorithm2e}[h]
  \DontPrintSemicolon
  \KwIn{$\{I_n^{(0)}\}_{n=1}^N$, binned indices; $\{C_n^{(1)}\}_{n=1}^N$, counts of bins of ideal histogram; $t$, linear parameter; {\tt max\_repeat}, maximum repeat parameter.}
  \tcp{$I_n^{(0)}$ is the list of indices in the $n$th bin. }
  $I^{(t)} \gets \left[\;\right]$ \tcp*{$I^{(t)}$ is an empty list.}
  \For{$n=1 : N$}{
  Compute $\alpha_n^{(t)}$.\tcp*{Use \cref{eq:newcounts,eq:alpha,eq:maxrepeat}.}
  $l \gets c_n^{(t)}$\\
  \If{$\alpha_n^{(t)}\geq1$}{
  Append $I_n^{(0)}$ \tt{floor}($\alpha_n^{(t)}$) times to $I^{(t)}$.\\
  $l \gets c_n^{(t)}-\left(\tt{count}(I_n^{(0)})\times \tt{floor}(\alpha_n^{(t)})\right)$. \\
  \tcp{$l$ is now an integer that satisfies $0 \leq l \leq \tt{count}(I_n^{(0)})$.}
  }
  Append a random subset of $I_n^{(0)}$ with length $l$ picked without replacement to $I^{(t)}$.
  }
  \Return $I^{(t)} $ \tcp*{New indices.}
  \caption{$I^{(t)}  \gets \tt{resample(}\{I_n^{(0)}\}_{n=1}^N, \{C_n^{(1)}\}_{n=1}^N, t\tt{)}$}
  \label{algo:sample_bins}
\end{algorithm2e}

\begin{algorithm2e}[h]
  \DontPrintSemicolon
  \KwIn{$\{I_n^{(0)}\}_{n=1}^N$, binned indices; $\{C_n^{(1)}\}_{n=1}^N$, counts of bins of ideal histogram; t, linear parameter; M, the size of dataset.}
  \tcp{$I_n^{(0)}$ is the list of indices in the $n$th bin. }
  $J \gets \tt{zeros(M)}$ \tcp*{$I^{(t)}$ is an empty list.}
  \For{$n=1 : N$}{
  Compute $\alpha_n^{(t)}$.\tcp*{Use \cref{eq:newcounts,eq:alpha}.}
  $J[I_n^{(t)}]=\alpha_n^{(t)}$
  }
  \Return $J $ \tcp*{New weights for samples.}
  \caption{$J \gets \tt{weights(}\{I_n^{(0)}\}_{n=1}^N, \{C_n^{(0)}\}_{n=1}^N, t, M\tt{)}$}
  \label{algo:weighted_loss}
\end{algorithm2e}

\section{Architecture Details} \label{sec:arch}
We process each of the input features separately with the 1D convolutions. 
To achieve this, we horizontally stack the features (vertical profiles of zonal wind, $U$, meridional wind, $V$, vertical wind, $\omega$, temperature, $T$ as ``channels''), resulting in a 2D input shape of {\tt nlev} $\times 4$.
(Note that the nomenclature of channels originates from Red Green Blue (RGB) channels in image processing.)
Additional information such as longitude, latitude, and surface pressure are concatenated to the flattened output of the encoder.
The resulting 1D array is pushed through dense layers intended to represent global relations.
Finally, the output from the dense section is reshaped to be processed via transposed convolutions and upsampling layers in the decoder.
\end{document}


%
%


\title{Supporting Information for ``Overcoming set imbalance in data driven parameterization: A case study of gravity wave momentum transport''}
%
%

%
%



\authors{L. Minah Yang \affil{1}, Edwin P. Gerber \affil{1}}


\affiliation{1}{Center for Atmosphere Ocean Science, Courant Institute of
Mathematical Sciences, New York University, New York, New
York, USA.}

%
%

%

\begin{article}

%
%

\noindent\textbf{Contents of this file}
\begin{enumerate}
\item Formal Algorithm Details
\end{enumerate}

\noindent\textbf{Additional Supporting Information (Files uploaded separately)}
\begin{enumerate}
\item Captions for Datasets S1 to Sx
\item Captions for large Tables S1 to Sx (if larger than 1 page, upload as separate excel file)
\item Captions for Movies S1 to Sx
\item Captions for Audio S1 to Sx
\end{enumerate}

\noindent\textbf{Introduction}


\noindent\textbf{Formal Algorithm Details}

\noindent\textbf{Data Set S1.} 


\noindent\textbf{Movie S1.} 


\noindent\textbf{Audio S1.} 
\end{article}
\clearpage